\documentclass[doublespacing]{elsart}

\usepackage{longtable}
\usepackage{rotating}
\usepackage[T1]{fontenc}

\usepackage{graphicx}

\usepackage{amssymb}
\providecommand{\tabularnewline}{\\}

\begin{document}

\begin{frontmatter}


 \title{Prototype scintillator cell for an In-based solar neutrino detector}
 \author{D. Motta\corauthref{cor}\thanksref{Saclay}}
 \ead{dario.motta@mpi-hd.mpg.de}
 \corauth[cor]{Corresponding author}
 \author{, C. Buck}
 \author{, F.X Hartmann}
 \author{, Th. Lasserre\thanksref{Saclay}}
 \author{, S. Sch\"onert}
 \author{, U.Schwan}
 \thanks[Saclay] {Now at CEA/Saclay, F-91191,
  Gif-sur-Yvette, France}
 \address{Max-Planck-Institut für Kernphysik (MPIK), Saupfercheckweg 1, 69117 Heidelberg, Germany}


\begin{abstract}

We describe the work carried out at MPIK to design, model, build and
characterize a prototype cell filled with a novel indium-loaded scintillator
of interest for real-time low energy solar neutrino spectroscopy.
First, light propagation in optical modules was studied with experiments
and Monte Carlo simulations. Subsequently a $5\,\textrm{cm}\times5\,\textrm{cm}\times100\,\textrm{cm}$
prototype detector was set-up and the optical performances of several
samples were measured. We first tested a benchmark PXE-based scintillator,
which performed an attenuation length of $\sim4.2\,\textrm{m}$ and
a photo-electron yield of $\sim730\, pe/\textrm{MeV}$. Then we measured
three In-loaded samples. At an In-loading of $44\,\textrm{g/l}$,
an energy resolution of $\sim11.6\,\%$ and a spatial resolution of
$\sim7\,\textrm{cm}$ were attained for $477\,\textrm{keV}$ recoil
electrons. The long-range attenuation length in the cell was $\sim1.3\,\textrm{m}$
and the estimated photo-electron yield $\sim200\, pe/\textrm{MeV}$.
Light attenuation and relative light output of all tested samples
could be reproduced reasonably well by MC. All optical properties
of this system have remained stable over a period of $>1\,\textrm{y}$.
\end{abstract}


\end{frontmatter}

\section{Introduction}

The use of indium as a target for the real-time spectroscopy of low
energy solar $\nu_{e}$ (pp, $^{7}$Be, CNO) has been thoroughly investigated
in the past \cite{Raju76,Pfeiffer-In,Payne-In,Suzuki-In} and reproposed
in recent years \cite{Frank-Stefan,Raju-In}. In the period 2001-2004
the feasibility study of an In-based solar neutrino detector has been
the focus of the LENS project \cite{LENS_gen}. The results presented here have
been carried out within the framework of this pilot phase.

The $\nu_{e}$ detection would proceed via an inverse-EC reaction
on $^{115}$In ($95.7\,\%$ isotopic abundance), followed by the delayed
$\gamma$ de-excitation of the daughter isomeric state $^{115}$Sn$^{*}$
$(612.8\,\textrm{keV},\,\tau=4.6\,\mu\textrm{s})$. The In target
is dissolved in an organic liquid scintillator at the highest concentration
compatible with adequate optical performances.

\begin{figure}
\begin{center}
 \includegraphics[width=1.0\columnwidth,
  keepaspectratio]{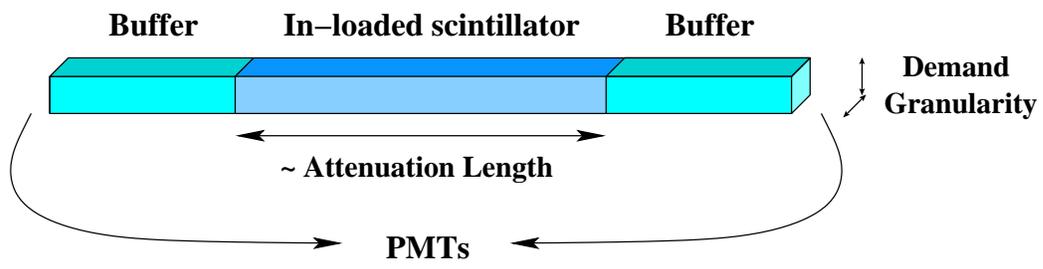}
\end{center}

\caption{{\small \label{fig:cell_gen_geo}Schematic layout of a generic In-loaded
scintillator cell.}}
\end{figure}

$^{115}$In decays $\beta$ with $\tau=6.4\times10^{14}\,\textrm{y}$,
which translates to a specific activity of $\simeq0.25\,\textrm{Bq/g}$
in $^{\textrm{nat}}$In. The suppression of the backgrounds associated
with the target activity requires a finely segmented detector. The
basic envisaged unit is an optical module, or \emph{cell}, filled
with In-loaded scintillator and viewed by two photomultiplier tubes
(PMTs), as shown in Fig. \ref{fig:cell_gen_geo}. %
The PMTs measure the energy deposited in the scintillator and, through
signal timing at both sides, the position of the event along the cell
longitudinal axis. There is no handle for the reconstruction of the
transversal coordinates, therefore the cell cross-section must be
chosen sufficiently small to control the background. The length is
upper-limited by light attenuation in the module. The range considered
for the cell dimensions is $5-10\,\textrm{cm}$ side, $\gtrsim1\,\textrm{m}$
length.

In order to test thoroughly the detector concept, a pilot experiment
\cite{LENS-Nu2004} has been set-up at the LENS Low Background Facility
(LLBF) \cite{LLBF}, at the Gran Sasso underground laboratory. The
pilot detector consists of a $3\times3$ matrix of $5\,\textrm{cm}\times5\,\textrm{cm}\times100\,\textrm{cm}$
scintillator cells. Four of the 9 cells are filled with In-loaded
liquid scintillators developed by us \cite{MPIK-Syn,MPIK-Lum,MPIK-BPO}
and another group of the LENS collaboration \cite{LNGS_scint}.

In this paper we describe the preparation, modeling, optimization
and final optical characterization of a prototype cell of an In-based
solar neutrino detector. In Sec. \ref{sec:Light-Piping} we report
on experimental studies on the light {}``piping'' in the cell. Sec.
\ref{sec:Scintillator} describes our In-loaded liquid scintillator,
with focus on its optical properties. Sec. \ref{sec:MC} is devoted
to the modeling of optics in the cell, and the results of preparatory
Monte Carlo studies are presented. In Sec. \ref{sec:Set-Up/Samples}
we describe the prototype setup and measurement campaigns. The results,
including the crucial test of the long-term ($>1\, y$) stability
of this system, are presented in Sec. \ref{sec:Results}. In Sec.
\ref{sec:Conclusions-Outlook} we give our conclusions and outlook.

\section{\label{sec:Light-Piping}Measurements of light piping}

The background suppression in an indium solar neutrino experiment
requires the unitary cell to have a small cross-section ($\sim5-10\,\textrm{cm}$
side). To limit the hardware costs, the unitary cell needs to be as
long as possible. In such a geometry (Fig. \ref{fig:cell_gen_geo}), photons  
undergo many reflections before hitting the PMTs and even a small
inefficiency can result in large light losses. These can degrade the
detector energy and spatial resolution, which are the fundamental
parameters toward background reduction. 

Denoting R as the surface reflectance, we have estimated that with
$\textrm{R}\lesssim98\,\%$ the original light intensity would drop
to less than 1/2 at $\sim2\,\textrm{m}$ distance, thus preventing
the construction of sufficiently long cells (\emph{i.e.}, of an affordable
detector). For the validation of the detection concept it was necessary
to check which level of piping efficiency can be met by using available
technology.

We have experimentally investigated the reflection coefficients of
two suitable piping mechanisms: total internal reflection (TIR) in
a quartz prototype cell; specular reflectance (SR) in a pipe lined
with a reflective foil. Quartz is the material chosen for the cells
of the final LENS pilot detector, due to its low activity and chemical
compatibility with organic solvents.

\subsection{Total internal reflection}

Total internal reflection from an optically denser to a less dense
medium provides in principle a $100\,\%$ efficient piping for all
impinging angles larger than $\theta_{c}=\arcsin(\textrm{n}{}_{o}/\textrm{n}{}_{i})$,
where n$_{i,o}$ are the refractive indices of the inner and outer
media. However, because of surface imperfections the
real reflectance is lower.

We have used a prototype quartz pipe to estimate the effective TIR
reflectance in a scintillation module. The pipe is $\simeq110\,\textrm{cm}$
long, its section is square with rounded corners, with $\simeq15\,\textrm{mm}$ 
external dimension and $\simeq2\,\textrm{mm}$ quartz thickness. 
Materials and surface finishing are similar to those of the final LENS 
prototype cells (to describe in Sec. \ref{sec:Set-Up/Samples}). 

\begin{figure}
\begin{center}\includegraphics[%
  width=1.0\columnwidth,
  keepaspectratio]{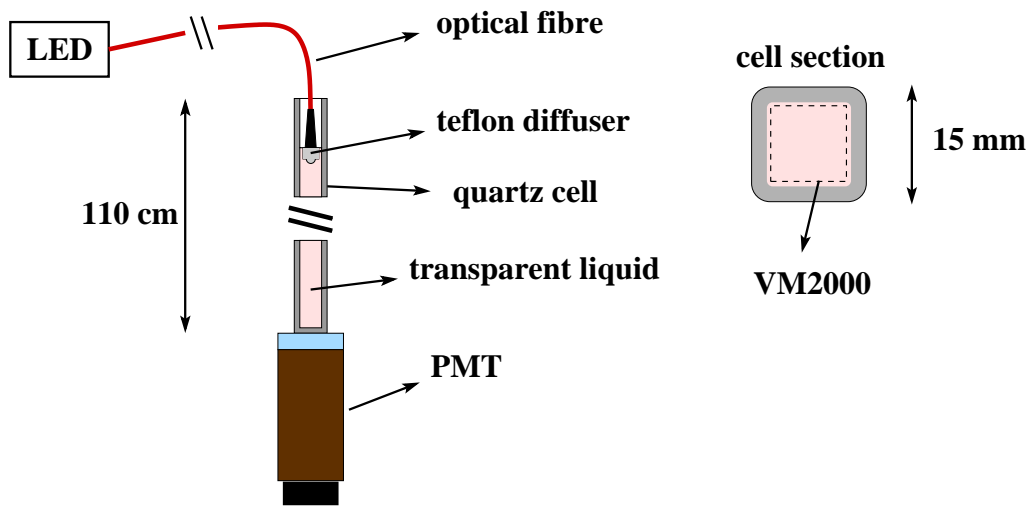}\end{center}

\caption{{\small \label{fig:piping-setup}Schematic view of the experimental
set-up for light piping measurements.}}
\end{figure}

A drawing of the experimental setup is shown in Fig. \ref{fig:piping-setup}.
The quartz tube is coupled to a PMT on one side and open on the other
side for filling and insertion of a light fiber, which is connected
to a $430\,\textrm{nm}$ LED%
\footnote{The results of similar measurements carried out at $380\,\textrm{nm}$
are reported in \cite{mythesis}%
}. The latter can be externally triggered to emit light in form of
fast pulses. At the free side of the fiber a Teflon cap diffuses the
LED light. %
 The cell was filled with liquids to simulate a scintillator-like
refractive medium. We have used {}``Uvasol'' cyclohexane ($\textrm{n}\simeq1.42$),
{}``Uvasol'' ethanol ($\textrm{n}\simeq1.37$) and {}``99+\%''
dodecane ($\textrm{n}\simeq1.42$), all showing excellent transparency
in the UV-visible range. The light fiber can be deployed at various
distances from the PMT and we measured the light intensity as a function
of the distance. 

\begin{figure}
\begin{center}\includegraphics[%
  width=1.0\columnwidth,
  keepaspectratio]{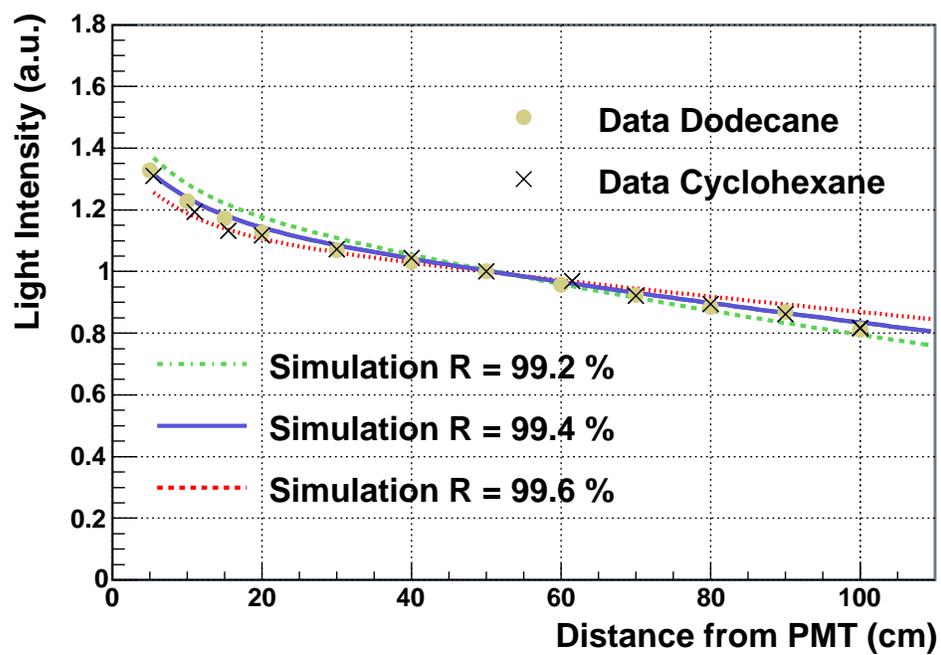}\end{center}

\caption{{\small \label{fig:TIR430}Measurement of light intensity as a function
of distance from the PMT. The cell was filled with cyclohexane and
dodecane. Three simulations for different TIR reflectance are reported.
Data and simulations are normalized to 1 at d = 50 cm.}}
\end{figure}

Fig. \ref{fig:TIR430} shows the results of the measurements and the
curves we predict for three values of the TIR reflectance. %
 The predictions are given by a photon-tracing MC simulation.

We observe that the two measurements using different filling liquids
are consistent with each other. The data analysis is based on the
comparison with the MC predictions (details in \cite{mythesis}).
The result is: \begin{equation}
\textrm{R}{}_{\textrm{TIR}}=(99.35\pm0.20)\%\label{eq:R-TIR}\end{equation}
The uncertainty ($1\sigma$) includes the statistical error ($\sim0.05\,\%$),
and the systematic errors due to the approximations in the MC ($\sim0.15\,\%$)
and to the uncertainty of the attenuation length of quartz and liquids
($\sim0.10\,\%$).

\subsection{Specular reflection by VM2000 foil}

TIR is a very efficient piping mechanism, however has the disadvantage
that a large fraction of the solid angle is not covered ($\sim50\,\%$,
for $\textrm{n}=1.5$ in the liquid). SR can be nearly equally efficient
at all angles, however the reflectance of conventional metal-coated
mirrors is limited. 

3M has recently issued the \emph{VM2000}, a multi-layer reflective
foil based on a novel technology \cite{VM2000}. We have measured
the absolute reflectance of different samples of VM2000 using the
\emph{V-W} accessory of our Varian \emph{Cary 400} spectrophotometer.

\begin{figure}
\begin{center}\includegraphics[%
  width=1.0\columnwidth,
  keepaspectratio]{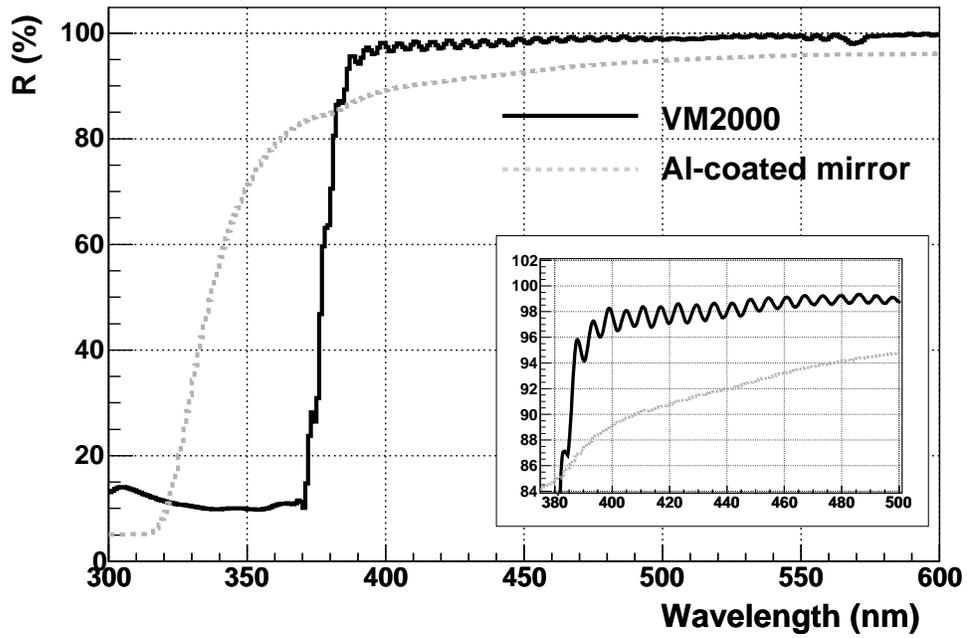}\end{center}

\caption{{\small \label{fig:VM2000_V-W}Spectral reflectance of a VM2000 sample.
For comparison, we also report the spectrum of an Al-coated mirror
employed in the H.E.S.S $\gamma$-ray Cherenkov telescope \cite{HESS},
as measured by us. The insert shows a magnification in the region
of interest.}}
\end{figure}

Fig. \ref{fig:VM2000_V-W} shows the result of a sample from a batch
shipped by 3M in 2002%
\footnote{Older foils had significantly different spectra and an overall worse
performance, see \cite{mythesis}.%
} and, for comparison, the reflectance of a conventional Al-coated
mirror. %
 We found that this foil has $\textrm{R}\gtrsim97\,\%$ for $\lambda\geq400\,\textrm{nm}$
($\textrm{R}\sim98\,\%$ at $430\,\textrm{nm}$). However, the systematic
error of the measurement ($\sim1\,\%$) is too large with respect
to the dependence of the response of an indium cell on \emph{}R. Moreover,
our equipment probes only one angle of incidence ($\theta=7^{\circ}$).

For these reasons, the system used to investigate TIR has been adapted
for the measurement of a SR-based cell: a long square profile of VM2000
was inserted into the quartz tube, so that the quartz internal surface
was lined with reflective foils (Fig. \ref{fig:piping-setup}). Data
were taken with the pipe filled with air and liquids to test whether
the optics of the foil is affected by the refractive index of the
coupled medium, and to cross check systematics due to the absorption
in the liquid. Data were analyzed by comparing with MC, and the result
at $430\,\textrm{nm}$ is: \begin{equation}
\textrm{R}_{\textrm{VM2000}}=(98.5\pm0.3)\%\label{eq:R-VM2000}\end{equation}
 for foil either in liquid and in air. The error ($1\sigma$) includes
the statistic uncertainty, the systematics deriving from the uncertainty
in the shape and dimensions of the pipe, and as well the systematics
associated with the MC-based analysis procedure. The result \ref{eq:R-VM2000}
is consistent with the V-W measurement at near-normal incidence, however
it is more precise and gives a reflectance averaged over the foil
surface and all angles of incidence.

\subsection{Implications}

Given the results of Eqs. \ref{eq:R-TIR} - \ref{eq:R-VM2000}, the
design optimizing light piping in an indium solar neutrino experiment
is a TIR-based cell externally wrapped with non-coupled VM2000 foil.

We have used the measured reflection coefficients to study the optics
of a cell via MC simulations. The results, published in \cite{mythesis},
indicate that with the aforementioned TIR/SR design, piping inefficiency
would not be limiting as far as the attenuation length due to absorption
in the scintillator is $\lesssim10\,\textrm{m}$.
This result was crucial to advance the LENS project in the pilot phase.

\section{\label{sec:Scintillator}In-loaded scintillator}

One approach initially explored in LENS for the production of In-loaded
scintillators was to dissolve In as a carboxylate compound $\textrm{InA}{}_{3}$
\cite{Raju-In}. However, it turned out that indium has a much higher
chemical reactivity to $\textrm{OH}{}^{-}$ than to the carboxylic
acids: $\textrm{In}{}^{3+}$ begins to react with water to form hydroxy
species even at pH 1 \cite{Frank}. 

Hydrolysis is a severe problem toward the synthesis of a stable and
robust In-loaded scintillator. The idea pursued at MPIK was to encapsulate
the In atom in a stable molecule that preserves its identity in the
organic solution. This was accomplished by using $\beta$-diketones
as ligand \cite{MPIK-Syn}.

Several In $\beta$-diketonates have been synthesized and studied.
The $\beta$-diketone finally chosen for the LENS prototype is the
simplest one, acetylacetone (Hacac), and the In complex is in the
form $\textrm{In(acac)}{}_{3}$. The latter is produced as a white
crystalline powder, purified by sublimation and finally dissolved
in the scintillator base. We found the solubility of $\textrm{In(acac)}{}_{3}$
in most of the standard organic scintillator solvents (including PC
and PXE) not to be sufficient for the demands of a neutrino detector.
However, anisol ensures a higher solubility, with a limit In-loading
of $7.9\,\%$ by weight \cite{MPIK-Syn}.

\subsection{Light yield}

\begin{figure}
\begin{center}\includegraphics[%
  width=1.0\columnwidth,
  keepaspectratio]{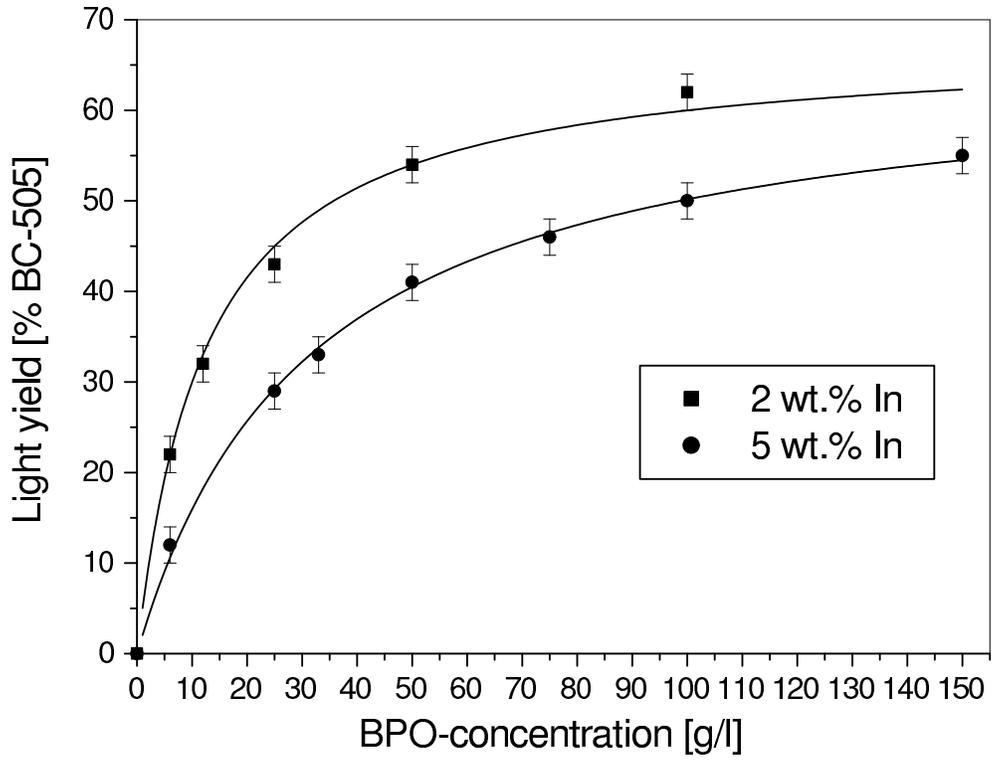}\end{center}

\caption{{\small \label{fig:LY-vs-BPO}LY versus BPO concentration in In-acac
scintillator samples having $\simeq20\,\textrm{g/l}$ and $\simeq50\,\textrm{g/l}$
In-loading. The figure shows our measurements (markers) and a fit
with a theoretical model (solid curves). The LY is given relative
to the BICRON BC505.}}
\end{figure}

The In(acac)$_{3}$ molecule has an absorption band that overlaps
with the UV emission of anisol \cite{MPIK-Lum}. In(acac)$_{3}$ gives
no detectable fluorescence when excited at its absorption band \cite{MPIK-Lum},
hence this molecule dissipates a part of the solvent excitation that
would otherwise convert into scintillation light. The strategy to
obtain good light yields (LY) is to enhance the {[}anisol $\rightarrow$
fluor{]} pathway relative to the quenching {[}anisol $\rightarrow$
In(acac)$_{3}${]} dissipation. The concentration of In(acac)$_{3}$
is constrained by the demand of high In-loading, therefore the only
handle for controlling the energy transfer is the choice of the fluor
and its concentration. We found that the best results are given by
BPO {[}2-(4-bi-phenyl)-5-phenyloxazole{]} \cite{MPIK-BPO,Buck} and
we observed that at high In-loading, the LY in a small sample increases
monotonically with the BPO concentration, up to its solubility limit.
The dependence of the LY on the fluor concentration is well explained
by theoretical models of energy transfer and light quenching \cite{MPIK-BPO,Buck}.
Fig. \ref{fig:LY-vs-BPO} shows the LY versus the BPO concentration
in two In-acac scintillator samples differing by In-loading. %
 At the benchmark In-loading of $50\,\textrm{g/l}$, notable light
yields are only obtained with fluor concentrations that are {}``unusual''
in standard organic scintillators, as regards the use in large scale
detectors.

\begin{figure}
\begin{center}\includegraphics[%
  width=1.0\columnwidth,
  keepaspectratio]{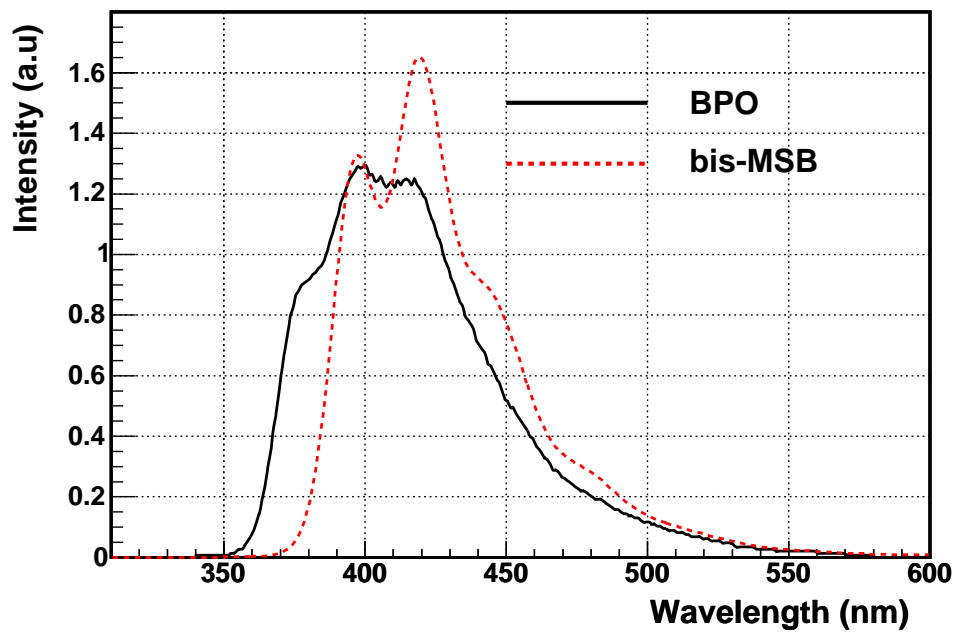}\end{center}

\caption{{\small \label{fig:em_spectra}Fluorescence spectra of BPO (at high
concentration) and bis-MSB, obtained via UV excitation at the wavelength
of maximum absorption.}}
\end{figure}

We also use bis-MSB {[}1,4-bis(2methyl-styryl)benzene{]}. The primary
LY in BPO systems is nearly independent of its concentration \cite{MPIK-BPO,Buck},
however bis-MSB shifts the scintillation light in a wavelength region
of higher transparency. Fig. \ref{fig:em_spectra} shows the emission
spectra of the two fluorescing molecular systems dissolved in our
scintillator.%

\subsection{\label{sub:Light-Attenuation}Light attenuation}

For application in a solar neutrino detector, the scintillator formulation
has to maximize the photo-electron yield (PY) in a cell. In this case,
light attenuation in the detector plays an important role.

\begin{figure}
\begin{center}\includegraphics[%
  width=1.0\columnwidth,
  keepaspectratio]{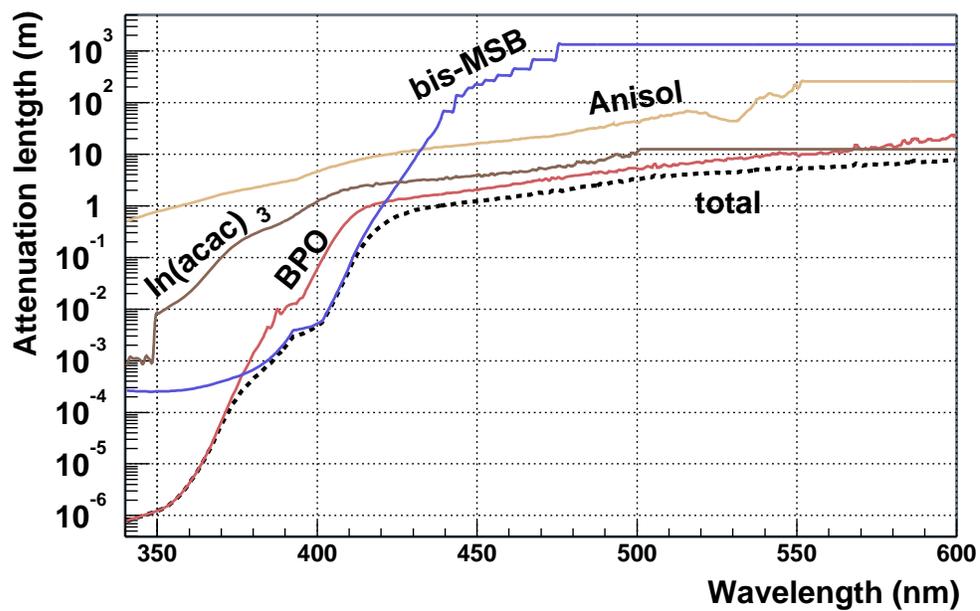}\end{center}

\caption{{\small \label{fig:att_length-Inscint}Attenuation lengths calculated
from spectrophotometric extinction data for {[}anisol / In(acac)$_{3}$
(In 50 g/l) / BPO (50 g/l) / bis-MSB (100 mg/l){]}.}}
\end{figure}

We have measured by spectrophotometry the extinction coefficients
of each scintillator component, as described in \cite{MPIK-Syn,MPIK-Lum,Buck}.
This information is combined with the concentration to calculate the
partial attenuation length, $\mu_{i}(\lambda)$, due to the $i^{th}$
solute. In our mixture the attenuation processes due to each component
are independent of each other, hence the resulting global attenuation
length $\mu$ is calculated as:\begin{equation}
\mu=\left(\sum_{i}\frac{1}{\mu_{i}}\right)^{-1}\label{eq:Lambert-II}\end{equation}
As an example, the partial $\mu_{i}$ and the total $\mu$ for the
system {[}anisol / In(acac)$_{3}$ (In 50 g/l) / BPO (50 g/l) / bis-MSB
(100 mg/l){]} are plotted in Fig. \ref{fig:att_length-Inscint}. %
 At the reference wavelength of $430\,\textrm{nm}$, the dominating
contribution is due to BPO $(\sim1.5\,\textrm{m})$. Fluorimetric
measurements have shown that BPO absorption for greater wavelengths
than $\sim410\,\textrm{nm}$ is not followed by re-emission and hence
results only in a light loss%
\footnote{Several attempts to purify BPO gave no significant improvement of
the transparency. %
} \cite{MPIK-BPO,Buck}. The purified In(acac)$_{3}$ contributes $\sim3\,\textrm{m}$,
anisol and bis-MSB both $\sim10\,\textrm{m}$. The resulting total
attenuation length at $430\,\textrm{nm}$ is $\mu\sim0.85\,\textrm{m}$.

The conclusion is that reasonable detector performances can be obtained
either with a more In-diluted scintillator (so that less fluor is
needed), or with shorter cells than initially envisaged. The decision
for the LENS pilot phase was the latter and hence the dimensions of
the prototype cells were fixed to $5\,\textrm{cm}\times5\,\textrm{cm}\times100\,\textrm{cm}$.
The focus has then shifted to find the parameter area where the best
compromise between primary LY and efficient light transport over $\sim1\,\textrm{m}$
distance is realized. This was accomplished by MC simulations.

\section{\label{sec:MC}Model of an In-loaded scintillator cell }

We have developed an optical model of the detector implementing a
detailed physical description of those features that we identified
as crucial. This model has been used to write a photon-tracing MC
of an In-loaded scintillator cell. 

\begin{figure}
\begin{center}\includegraphics[  
  width=1.0\columnwidth,
  keepaspectratio] {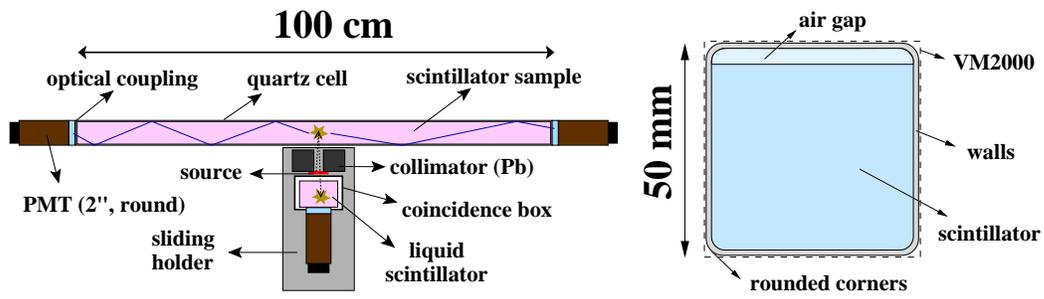}\end{center}

\caption{{\small \label{fig:prototype-CBS-Sec}Left: schematic view of our
experimental set-up for the optical measurement of the prototype cell. }}

{\small Right: section of the cell. All shown geometrical features
are implemented in the MC simulations.}
\end{figure}

The MC implements a geometry that faithfully reproduces the actual
In prototype cell, to be described later (shown in Fig. \ref{fig:prototype-CBS-Sec}).
The scintillator formulation is based on the aforementioned anisol/In(acac)$_{3}$/BPO/bis-MSB
mixture. We made the assumption that light is emitted according to
the fluorescence spectrum of the last non-radiatively excited component.
We have determined the nature of the donor-acceptor energy transfer
as discussed in \cite{MPIK-Lum,Buck}. We found that, for our concentration
levels, the transfer anisol $\rightarrow$ BPO is non-radiative, while
BPO $\rightarrow$ bis-MSB is predominantly radiative. As a consequence,
we assume the BPO emission spectrum as the primary fluorescence (Fig.
\ref{fig:em_spectra}).

The basic interactions of optical photons in a medium are \emph{absorption},
\emph{elastic scattering} and \emph{inelastic scattering}. In the
latter case the photon is absorbed by a fluorescing molecule, which
de-excites emitting a photon of longer wavelength. In our systems
different chemical species are present and each one contributes to
some or all of the above processes. At the spectrophotometer, absorption,
elastic and inelastic scattering are indistinguishable, as they all
result in removing photons from a collimated beam. Therefore, it is
necessary to evaluate independently their relative probabilities.
We have neglected elastic scattering, since it is sub-dominant in
most cases. Consequently, the measured extinction is considered as
the sum of absorption and inelastic scattering. A fluorescence quantum
yield, $\phi$, can be defined as the probability that an absorption
is followed by a prompt re-emission. In Table \ref{tab:fluoresc-param}
the values of $\phi$ used in our model are reported. %
 We assume $\phi=0\,\%$ for anisol at all wavelengths because its
emission spectrum has no overlap with the emission of BPO \cite{MPIK-Lum}.
 BPO is expected to have very high quantum yield, as the
literature value for the similar PPO is $\phi\simeq100\,\%$ \cite{fluors QE}.
However, we chose a more conservative $90\,\%$, with cut-off at $410\,\textrm{nm}$
(see Sec. \ref{sub:Light-Attenuation}). For bis-MSB we use the literature
$\phi$ \cite{fluors QE} and our fluorimetric measurements for the
cut-off \cite{Buck}. $\textrm{In(acac)}{}_{3}$ and quartz do not
show fluorescence in the visible, hence $\phi=0$.

In our MC simulation all components interact with light independently
of each other. The respective attenuation lengths are calculated from
the concentrations and extinction coefficients, as in Fig. \ref{fig:att_length-Inscint}.
The parameters in Table \ref{tab:fluoresc-param} are used to decide
whether an interaction is an absorption or a wavelength shift.
Light piping in the cell is simulated via TIR and SR using Eqs. \ref{eq:R-TIR}
and \ref{eq:R-VM2000}. We fix the refractive index of scintillator, 
quartz and PMT window to 1.5.
The PMTs are coupled to the cell ends with $\pi/4$ surface coverage
(a disk inscribed in a square). For their sensitivity, the typical
{}``green-enhanced'' bialkali QE is chosen, and the values are downscaled
of a factor 0.75 to take into account the reduced sensitive area and
the collection efficiency. Light interaction with the PMTs is modeled
as in \cite{PMT_optics}.

The output of the MC is the photo-detection probability, which can
be converted in absolute units by folding with the scintillator LY.

\subsection{\label{sub:Scintillator-Optimization}Scintillator optimization}

One crucial question addressed by MC simulation is the determination
of the optimal BPO concentration, for the given $50\,\textrm{g/l}$
In-loading. For this analysis the bis-MSB concentration was fixed
to $100\,\textrm{mg/l}$, and BPO was varied. \emph{}

\begin{figure}
\begin{center}\includegraphics[%
  width=1.0\columnwidth,
  keepaspectratio]{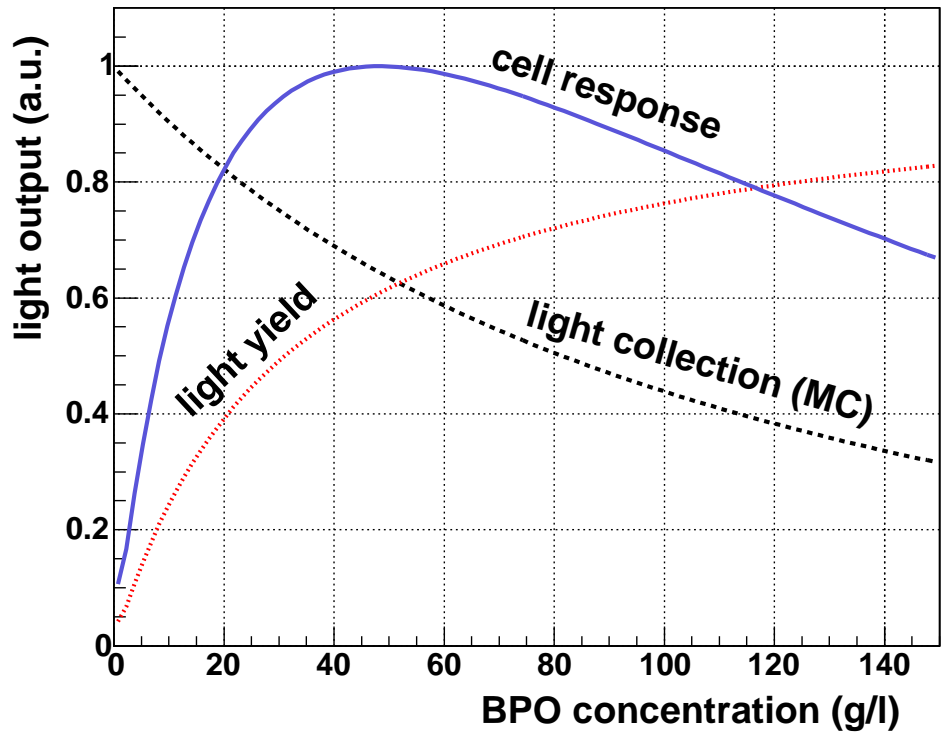}\end{center}

\caption{{\small \label{fig:Light_vs_BPO}Effect of the variation of the BPO
concentration on the detector response. The latter is the product
of the scintillator LY (Fig. \ref{fig:LY-vs-BPO}) times the light
collection. All functions are normalized to 1 at their maximum.}}
\end{figure}

Taking as reference parameter the light output for a source at the
cell center, Fig. \ref{fig:Light_vs_BPO} illustrates the competition
between LY and light transport. The best compromise is expected at 
$\sim50\,\textrm{g/l}$ BPO, which gives $\textrm{LY}\sim42\,\%$
relative to BC505%
\footnote{\label{foot:AbsoluteLY}The BICRON specification for BC505 is $80\,\%$
of Anthracene. The BC505 measured by us is $\sim105\,\%$ of the Borexino
mixture \cite{Borexino}, reported as giving $(11500\pm1000)\, pe/\textrm{MeV}$
\cite{Borexino-scint}.%
}. 

\begin{figure}
\begin{center}\includegraphics[%
  width=1.0\columnwidth,
  keepaspectratio]{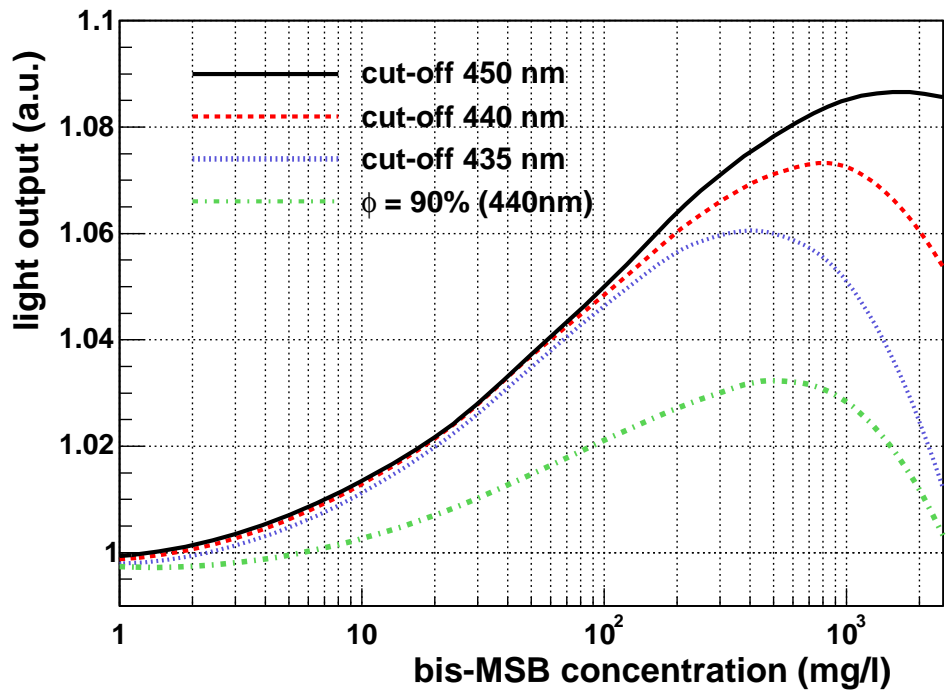}\end{center}

\caption{{\small \label{fig:Light_vs_Bis-MSB}MC-simulated effect of the variation
of the bis-MSB concentration on the detector light output. The predictions
for 3 re-emission cut-offs are shown ($\phi=94\,\%$) and for $\phi=90\,\%$
(cut-off at $440\,\textrm{nm}$). All curves are normalized to the
case of no bis-MSB.}}
\end{figure}

In a similar way, the effect of the variation of the bis-MSB concentration
was studied via MC simulation. The results are shown in Fig. \ref{fig:Light_vs_Bis-MSB},
for a fixed BPO concentration of $50\,\textrm{g/l}$ and source at
the cell center. The MC predictions are strongly model-dependent and
the optimal bis-MSB concentration shifts to higher values along with
the cut-off for re-emission and with $\phi$. %
 Using the best values from the literature and our measurements (Table
\ref{tab:fluoresc-param}), the optimal concentration is expected
in the few $10^{2}\,\textrm{mg/l}$ to $1\,\textrm{g/l}$ range, and
should yield a $\sim6-8\,\%$ relative increase of the light output
with respect to the case of no bis-MSB. The bis-MSB concentration
is therefore predicted to be a non-critical parameter. This is due
to the fact that the BPO emission is already peaked at long wavelengths
(Fig. \ref{fig:em_spectra}) and that BPO can itself act as a wavelength
shifter (WLS) in the range $\lambda\lesssim410\,\textrm{nm}$.

\section{\label{sec:Set-Up/Samples}Prototype set-up and measurement campaigns}

The prototype cell reported here is one of a set of 12 identical quartz
modules built by MPIK for the LENS pilot experiment \cite{LENS-Nu2004}.
The raw profile tubes have been manufactured by the \emph{Heraeus
Quartz Glas GmbH} out of {}``synthetic quartz crystals''. \emph{Helma
GmbH} has welded the two side windows and manufactured two openings
and stoppers for liquid filling/removal. The cells have been finally
annealed at Heraeus. 

The cell dimensions are $5\,\textrm{cm}\times5\,\textrm{cm}\times100\,\textrm{cm}$,
the section is square with rounding at the corners ($4\,\textrm{mm}$
radius of curvature) and the wall thickness is $2\,\textrm{mm}$.
A $5\,\textrm{mm}$ air gap is left on top of the scintillator to
avoid that this touches the stoppers of the filling openings.

\subsection{Experimental technique}

A system has been set up to measure the optical performances of the
prototype cell, including light attenuation, energy and spatial resolution,
PY. In order to test the detector response to the deposition of single
known energies, the Compton-back-scattering (CBS) technique was implemented.

Fig. \ref{fig:prototype-CBS-Sec} shows a drawing of the experimental
set-up. The cell is irradiated with a collimated $\gamma$ source
and the light pulses resulting from the energy deposited by the scattered
electrons are measured in coincidence on both PMTs. An additional
liquid scintillation unit is located on the back side of source and
collimator, opposite to the cell. A three-fold coincidence can be
set, which selects the events with simultaneous energy deposition
in the cell and in the back-scattering box.

\begin{figure}
\begin{center}\includegraphics[%
  width=1.0\columnwidth,
  keepaspectratio]{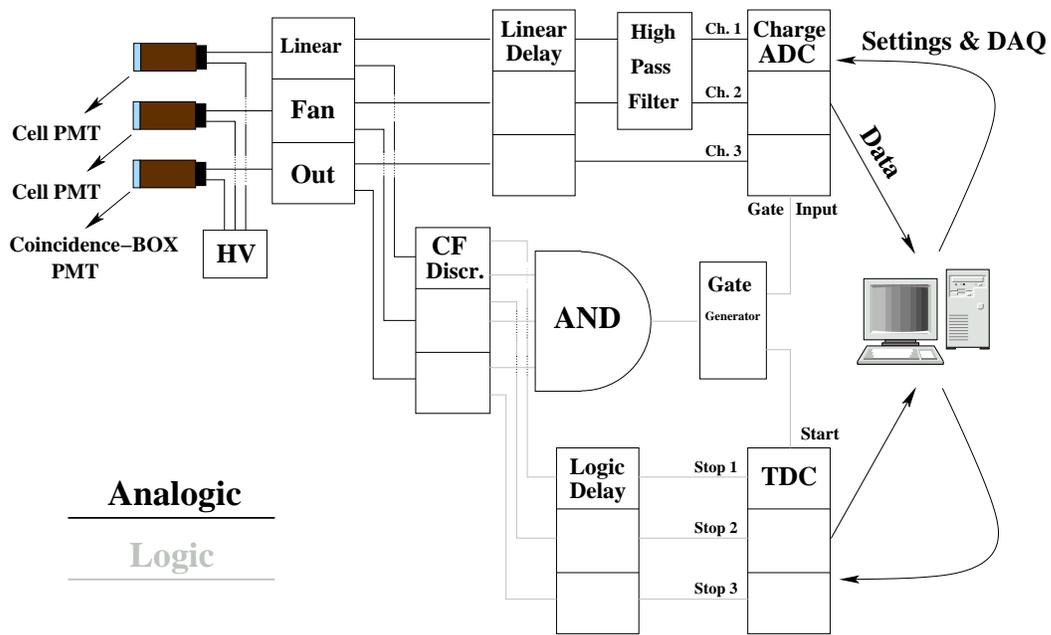}\end{center}

\caption{{\small \label{fig:Electronics-Dark-Room}Logic schema of the electronics.}}
\end{figure}

The logic scheme of the electronics is shown in Fig. \ref{fig:Electronics-Dark-Room}.
The PMTs used are the ETL9954B. Their analogical signals are doubled;
one copy is discriminated and then used to form the trigger condition
and the logic \emph{stop} signals for a Time-to-Digital-Converter
(TDC); the other copy is delayed of the time required for the system
to trigger and then sent to an integrating charge-sensitive Analog-to-Digital-Converter
(ADC).

In order to improve the S/B ratio of the CBS measurement, we apply
event-by-event cuts. A large part of the backgrounds is removed by
requiring that the energy in the coincidence box is $E\leq E_{\gamma}-E_{CBS}$
and that the TDC spatial reconstruction of the event is consistent
with the position of the irradiated volume. 

Two $\gamma$ sources have been utilized: $^{137}$Cs and $^{54}$Mn,
which deliver CBS recoil electrons of $477\,\textrm{keV}$ and $639\,\textrm{keV}$,
respectively. These two energies are particularly significant for
an In detector: the former is close to the energy of the second $\gamma$
of the $\nu$-tag ($498\,\textrm{keV}$) and is representative of
the \emph{hard-Bremsstrahlung} background \cite{LENS_gen}. The CBS
of the $^{54}$Mn source has an energy similar to the total energy
released by the $\nu$-tag ($613\,\textrm{keV}$). With these two
sources it is possible to test directly how well the $\nu$-tag can
be resolved from background events in the range of the $^{115}$In
$\beta$ end-point energy. This is one of the most crucial issues
for the feasibility of an indium solar neutrino experiment \cite{LENS_gen}.

\subsection{Measurements}

We have measured 4 scintillator samples, whose characteristics are
reported in Table \ref{tab:samples}. We will refer to them using
the same nomenclature in the table. After each cell filling the scintillator
has been flushed with nitrogen for about one hour to sweep oxygen,
which can decrease the LY and lead to chemical degradation. %

\subsubsection{Benchmark PXE cell}

The PXE-based scintillator serves to study the detector performance
in a benchmark case. The PXE comes from a batch used by Borexino in
the CTF \cite{PXE-paper}. We measured the LY to be $(83\pm4)\%$
relative to BC505. The sample has been purified by us using a solid
(neutral pH silica-gel) column plant as reported in \cite{PXE-paper}.
The resulting attenuation lengths at various wavelengths are given
in the first row of Table \ref{tab:samples_att-lengths}. In the range
of the emission peak $\mu\gg1\,\textrm{m}$, therefore the measurement
of this sample tests the optical integrity of the quartz cell and
cross-checks the reflection coefficients given in Sec. \ref{sec:Light-Piping},
through comparison with the MC. Furthermore, assuming that no highly
In-loaded scintillator would ever perform better than this sample,
the PXE performances provide an upper limit for those of an In-cell
in the considered design.

\subsubsection{In-loaded cells}

We have synthesized and purified several hundreds grams of In(acac)$_{3}$
crystals for use in these optical measurements and for loading in
the final LENS pilot detector at Gran Sasso. The three scintillator
samples In-l, In-h1 and In-h2 have been prepared by subsequent additions
of components to a common base. The measurement of the In-l sample
aims at studying the cell performance in the lowest In-loading range
that is considered of interest for a solar neutrino detector. The
samples In-h1 and In-h2 have In-loading in the range targeted by LENS
and are the base case for the pilot phase. Their BPO concentration
was chosen based on the results of the MC studies presented in Sec.
\ref{sub:Scintillator-Optimization} and they only differ by the bis-MSB
concentration. Table \ref{tab:samples_att-lengths} reports the attenuation
lengths of the three samples at various wavelengths. %

\section{\label{sec:Results}Results}

\subsection{\label{sub:PXE}Benchmark PXE cell}

Two sets of measurements have been performed:

\begin{enumerate}
\item TIR light piping
\item TIR plus VM2000 light piping 
\end{enumerate}
For the measurements 1 light was piped via TIR at the quartz/air and
scintillator/air boundaries. For the second campaign a non-coupled
VM2000 profile has been wrapped around the cell.

\subsubsection{\label{sub:PXE_att}Light attenuation}

\begin{figure}
\begin{center}\includegraphics[%
  width=1.0\columnwidth,
  keepaspectratio]{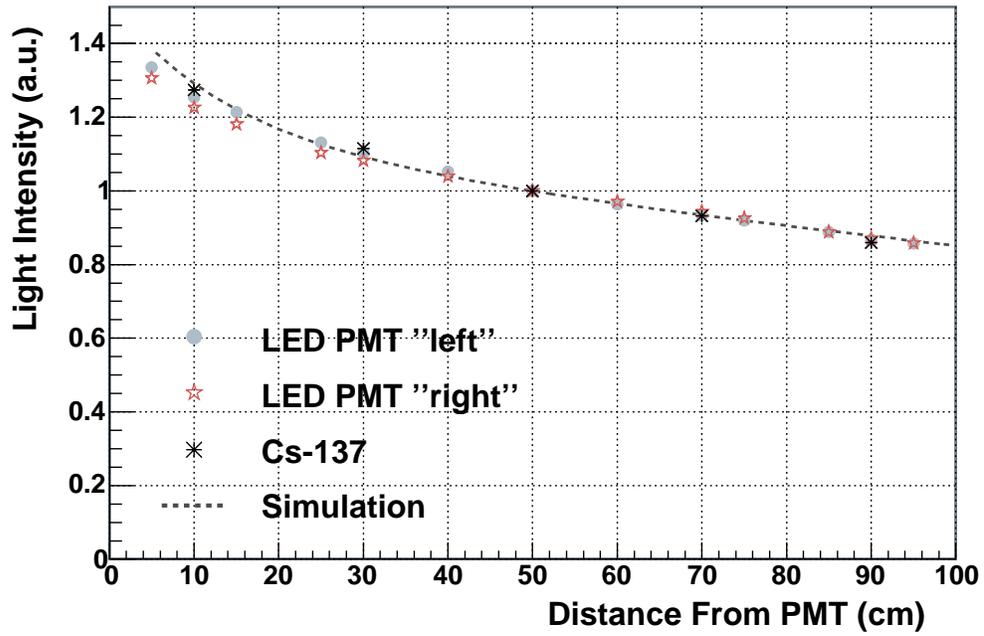}\end{center}

\caption{{\small \label{fig:pxe_tir_att}Light intensity versus source distance
for the PXE sample with TIR piping. The estimated error of the data
points is $\sim1\%$. Data and simulation are normalized to 1 at $\textrm{d}=50\,\textrm{cm}$. }}
\end{figure}

With the TIR cell the light attenuation curve has been measured by
scanning the detector response to a $380\,\textrm{nm}$ collimated
LED. The LED excites the scintillator fluors, which re-emit light
isotropically. The resulting light attenuation curves are
shown in Fig. \ref{fig:pxe_tir_att}. For comparison, the graphics
also reports the data-points obtained with the CBS peak of the $^{137}$Cs
source (measurement to be described later). %
 The prediction of the MC simulation of the system is also shown in
the figure.

No significant differences are observed in the attenuation curves
measured by the two PMTs, as expected, and the CBS data are consistent
with the LED measurement. A small departure between simulation and
experiment is observed at short distances, where the simulation predicts
a slightly faster attenuation%
\footnote{Since we compare the attenuation curves with each other upon normalization
at the cell center, the fact that the simulation stays above the data
at short distance simply means that light is expected to be attenuated
faster in this range.%
}. However, in the rest of the curve MC and data are in good agreement.

\begin{figure}
\begin{center}\includegraphics[%
  width=1.0\columnwidth,
  keepaspectratio]{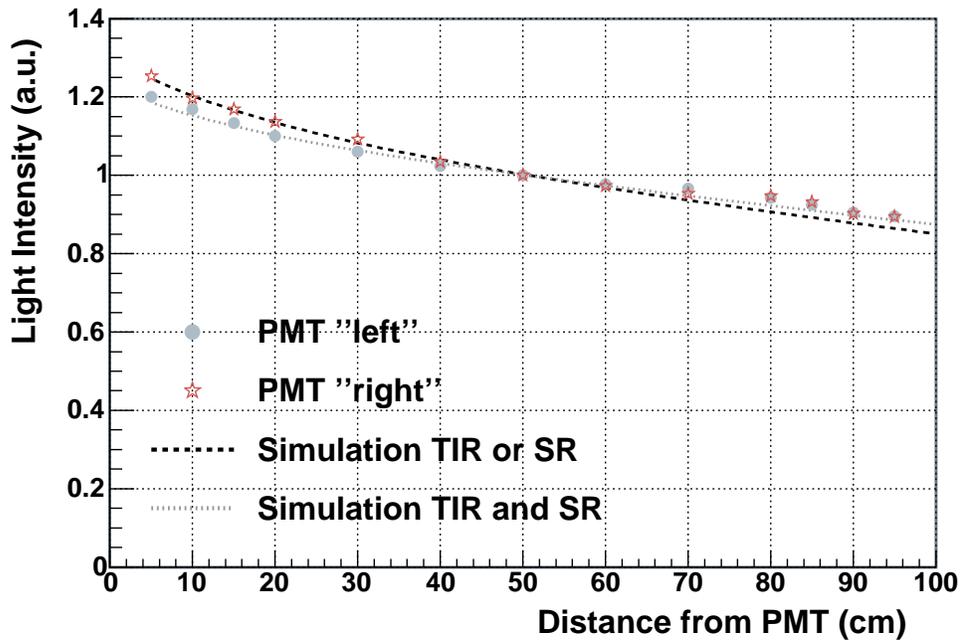}\end{center}

\caption{{\small \label{fig:pxe_foils_att}Light attenuation curves for the
PXE sample, TIR piping plus VM2000 wrapping. The markers represent
the data, the curves are simulations of the system for the two limit
cases discussed in the text. The estimated error of the data points
is $\sim2.5\,\%$. Data and simulations are normalized to 1 at $\textrm{d}=50\,\textrm{cm}$.}}
\end{figure}

For the cell wrapped with reflective foil, the light
attenuation curve could not be measured with the external LED, since
the transparency of the VM2000 is very poor. The cell has been irradiated
with the collimated $^{137}$Cs source and the system operated in simple two-fold
coincidence mode, to select all Compton interactions
in the cell. The position of the Compton-edge was used to estimate the relative
light intensity. The resulting attenuation curves are shown in
Fig. \ref{fig:pxe_foils_att}. %

The measurements of the two PMTs are in slight disagreement in the
initial part of the curves, probably due to a slow drift of the gain 
of one PMT. There is however good agreement for $\textrm{d}>40\,\textrm{cm}$.

The predictions of two MC simulations are also reported in the figure.
Two limit cases are shown, which are defined {}``TIR or SR'' and
{}``TIR and SR''. In the former, light reflects either via TIR or
SR, depending whether $\theta>\theta_{c}$ or $\theta<\theta_{c}$.
{}``TIR and SR'' assumes that all the light lost by TIR at $\theta>\theta_{c}$
has the opportunity to be reflected by the foils. This results in
an effective $\textrm{R}\sim99.99\,\%$ for $\theta>\theta_{c}$.
Light diffusion at the interfaces is not considered.

We fitted the attenuation curves with a double exponential function:\begin{equation}
f=I_{s}e^{-(x-x_{0})/\mu_{s}}+I_{l}e^{-(x-x_{0})/\mu_{l}}\label{eq:Light_Att_fitf}\end{equation}
where $x_{0}$ is the distance of the closest data point (here $x_{0}=5\,\textrm{cm}$).
The fit returns two effective attenuation lengths ($\mu_{s}$ and $\mu_{l}$), the
weights at the reference distance $x_{0}$ ($I_{s}$ and $I_{l}$)
and the uncertainties for those parameters. The ratio $W_{s,l}=I_{s,l}/(I_{s}+I_{l})$
gives the fraction of the intensity measured by the PMT at $x_{0}$ that is
attenuated with a $\mu_{s,l}$ attenuation length. Most important
for this study is the longer attenuation component, $\mu_{l}$, which
characterizes light attenuation at a distance from the PMT where geometrical
effects and short-range light-loss mechanisms have decayed out. 

The analysis of the experimental and simulated curves
of Figs. \ref{fig:pxe_tir_att} and \ref{fig:pxe_foils_att} is given
in Table \ref{tab:pxe_att-fitpar}. %
 In the case of TIR piping, about $(85-90)\%$ of the light detected
from $5\,\textrm{cm}$ distance propagates in the cell with an effective
attenuation length of $\sim3\,\textrm{m}$. The simulation predicts similar
parameters. This means that the physics of light propagation in the cell 
is well understood and no light loss mechanisms beside the known ones 
are present.

For the cell wrapped with VM2000 foil $\mu_{l}=(4.2\pm0.3)\,\textrm{m}$
with weight $\sim91\,\%$, significantly better than the performance
of TIR alone. This result disfavors the {}``TIR or SR'' scenario:
for this case, the effective attenuation length should not improve,
as the VM2000 would simply pipe light outside the TIR acceptance,
however with a lower efficiency than TIR (cf. Eqs. \ref{eq:R-TIR}
and \ref{eq:R-VM2000}).

The longitudinal scan of the cell response allows calibrating the
TDC. The channel-to-position scale is linear with good approximation.
From the width of the TDC distributions we deduce that a single LED
event originating $\sim10^{3}$ photo-electrons (estimated from the width
of the ADC distribution) can be reconstructed with a spatial resolution
of $\sim2\,\textrm{cm}\,(1\sigma)$.

\subsubsection{\label{sub:PXE_CBS}CBS measurements}

\begin{figure}
\begin{center}\includegraphics[%
  width=1.0\columnwidth,
  keepaspectratio]{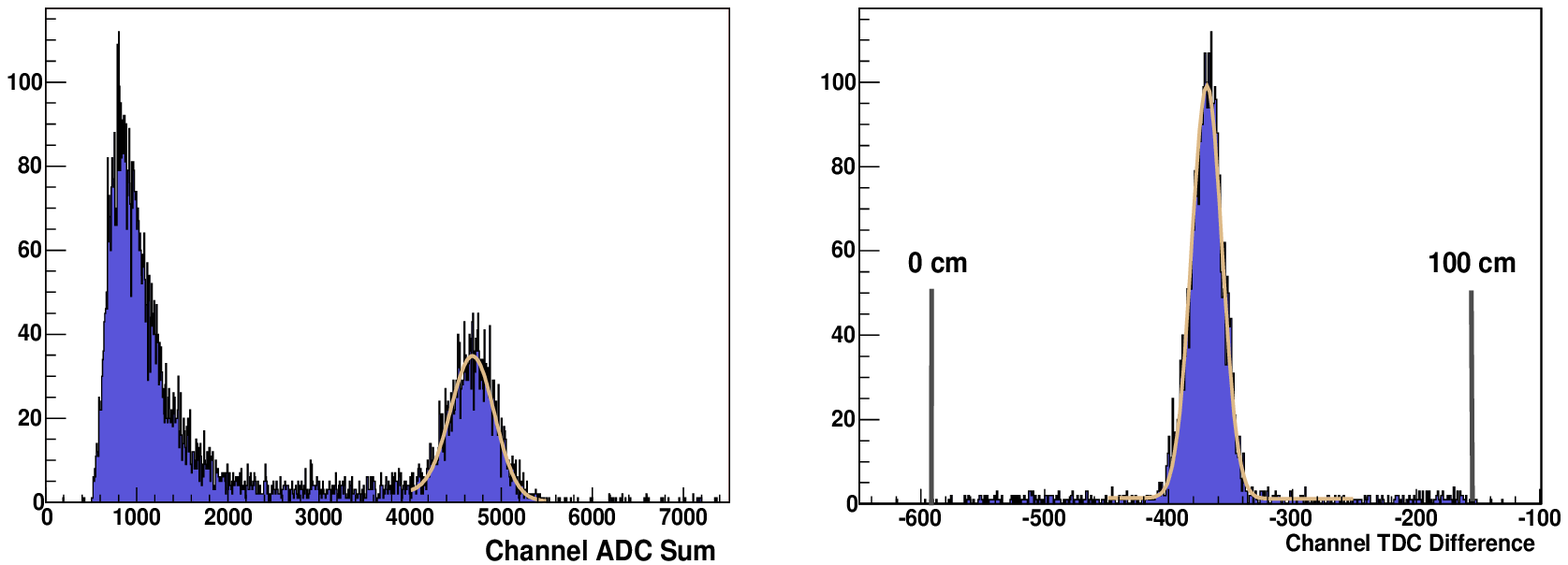}\end{center}

\caption{{\small \label{fig:pxefoils-ADC-TDC}CBS spectra for the PXE cell
with VM2000, $^{137}$Cs source at cell center.}}

{\small Left: distribution of the total charge, calculated by summing
the ADC values of the two PMTs.}{\small \par}

{\small Right: distribution of the time difference between the two
PMTs for the events around the CBS peak. The TDC signals corresponding
to the ends of the cell are indicated by the two vertical segments.}
\end{figure}

The ADC and TDC spectra of a $^{137}$Cs CBS run for the cell wrapped
with VM2000 are shown in Fig. \ref{fig:pxefoils-ADC-TDC}. %
 The energy spectrum after cuts shows a nearly
Gaussian CBS peak, with a modest background contamination on the left
side of the peak. The latter is sufficiently small to allow a precise
determination of the energy resolution by fitting with a Gaussian
plus an asymmetric background. The TDC spectrum obtained by selecting
only the events in the CBS energy window has a prominent Gaussian
peak corresponding to the source position. 

The results of the analysis of ADC and TDC CBS data for both measurements
of the PXE sample are reported in Table \ref{tab:pxe_CBS-analysis}. %
 At 477 keV, the $1\sigma$ energy resolution is $(8.8\pm0.2)\%$ for
TIR piping, $(5.4\pm0.2)\%$ for mixed TIR/SR piping, nearly constant
over the cell length. At the same energy, the $1\sigma$ spatial
resolution is $(4.4\pm0.2)\,\textrm{cm}$ and $(3.2\pm0.2)\,\textrm{cm}$
for the TIR and TIR/SR cells, respectively, source at center; it gets
slightly worse at the cell ends.

\subsubsection{Photo-electron yield}

We will consider first the measurement of the TIR cell. The detector
energy resolution gives an indication of the PY. If it is assumed
that the width of the CBS peak is dominated by the statistical fluctuations
in the average total number of \emph{pe}, $\bar{N}_{s}$, we find
$\bar{N}_{s}=123\pm5$ at $477\, \textrm{keV}$, source at center. This would
translate to $\textrm{PY}=(260\pm10)$ \emph{pe}/MeV. Repeating the
calculation for each PMT leads to $\bar{N}_{l}=57\pm2$ and $\bar{N}_{r}=71\pm2$.

However, the energy resolution is not expected to be purely statistical.
In order to estimate independently the PY, we have calibrated the
system by measuring the response to a \emph{spe}. For this calibration
the LED has been used. The light intensity was tuned to a sub-Poisson
regime ($pe/\textrm{pulse}\ll1$), so that the probability to measure
events with more than one \emph{pe} is negligible. The trigger was
synchronized with the external LED trigger signal and a $\times50(1\pm0.03)$
amplification stage was introduced.

The result of the calibration is: $N_{l}=58\pm9$
and $N_{r}=73\pm3$. The calibration of the {}``right'' PMT is more precise due
to a better peak/valley ratio of the \emph{spe} response.

The estimation of the number of \emph{pe} based on this calibration
is consistent with the energy resolution. Therefore we conclude that
in this regime the detector resolution is dominated by the statistical
fluctuations of the number of \emph{pe}.

For the TIR/SR cell, the energy resolutions of the $^{137}$Cs and
$^{54}$Mn CBS lines correspond to a limit $\textrm{PY}=(720\pm25)\, pe/\textrm{MeV}$
and $\textrm{PY}=(740\pm30)\, pe/\textrm{MeV}$, respectively. The
\emph{spe} calibration gives an average of $(171\pm7)\, pe/\textrm{PMT}$
at $477\,\textrm{keV}$, translating to $\textrm{PY=}(715\pm30)\, pe/\textrm{MeV}$.
Therefore energy resolution and \emph{spe} calibration are also in
this case in agreement with each other.

Table \ref{tab:PXE_PY} summarizes the analysis of the PY based on
the CBS measurements and the \emph{spe} calibration. The results of the MC
are also reported. The MC predicts more than twice as much light as measured.
This discrepancy must be understood, since the feasibility of an In experiment
depends dramatically on the detector energy resolution. This question will
be addressed again in Sec. \ref{sub:Summary-and-Discussion}.

We observe that the use of the VM2000 foil improves the light collection of more
than a factor 2. This shows that the foil serves not only to
pipe the light outside the TIR acceptance, but also compensates the
non ideal geometry of the quartz cell (rounded corners) and other
{}``TIR leakages''. This effect is well described by our model: the 
MC predicts $\textrm{PY}{}_{\textrm{TIR/SR}}/\textrm{PY}_{\textrm{TIR}}$
$\sim2.3-2.5$, depending on the assumed scenario. The experimental
value is $\sim2.4$.

\subsection{\label{sub:Indium}Indium-loaded cells}

The measurements with PXE have shown that the prototype cell works
properly and that the VM2000 wrapping increases the PY of a factor
of $\sim2.5$. Consequently this design was also chosen for the indium
measurements.

\subsubsection{Light attenuation}

Due to the lower energy resolution, the Compton
edge smears into a shoulder and no clear peak is observed. For this reason, we
preferred to perform the scan of the cell with CBS measurements.

\begin{figure}
\begin{center}\includegraphics[%
  width=1.0\columnwidth,
  keepaspectratio]{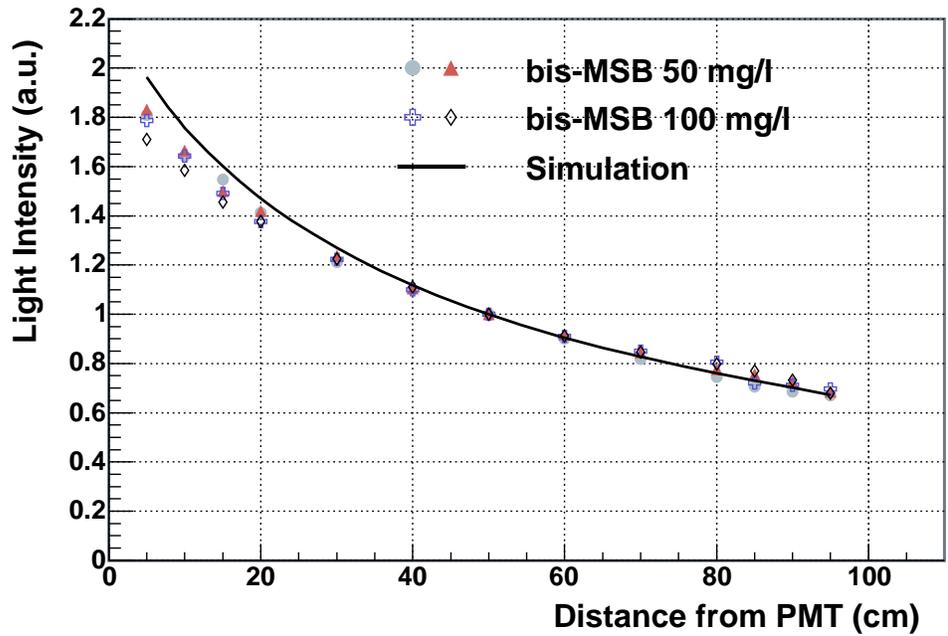}\end{center}

\caption{{\small \label{fig:In-high_att}Light attenuation curves for the
In-h1 and In-h2 samples. The markers represent the data, the solid
line is a simulation of the system. Only one simulation is shown,
because the MC predicts no significant differences for the two samples.
The estimated error of the data points is $\sim2.5\,\%$. Data and
simulation are normalized to 1 at $\textrm{d}=50\,\textrm{cm}$.}}
\end{figure}

The results for the In-h1 and In-h2 samples are shown
in Fig. \ref{fig:In-high_att} together with the MC-predicted light
attenuation curve. %
 No significant differences are observed in the attenuation of the
two samples. The simulation reproduces very well the long-range behavior
of the experimental curves, whereas it predicts a slightly faster
light attenuation at $\textrm{d}\lesssim30\,\textrm{cm}$. This means
that this detector performs better than expected at short distance,
and as expected at long distance. 

The results of the fit of experimental and simulated curves for all
In samples are given in Table \ref{tab:In_att-fitpar}. %
 The high-loading samples give $\mu_{l}\simeq1.3\,\textrm{m}$ for
$75-80\,\%$ of the light. The experimental finding that they exhibit
very similar attenuation is also predicted by the MC. The longer $\mu_{l}$
of the simulations is misleading: it simply results from the smaller
weight of the long-range component ($\mu_{l}$ and $W_{l}$ anti-correlate
in the fit). The In-l sample has a slightly better performance, as
expected.

\subsubsection{CBS measurements}

\begin{figure}
\begin{center}\includegraphics[%
  width=14cm,
  height=5cm]{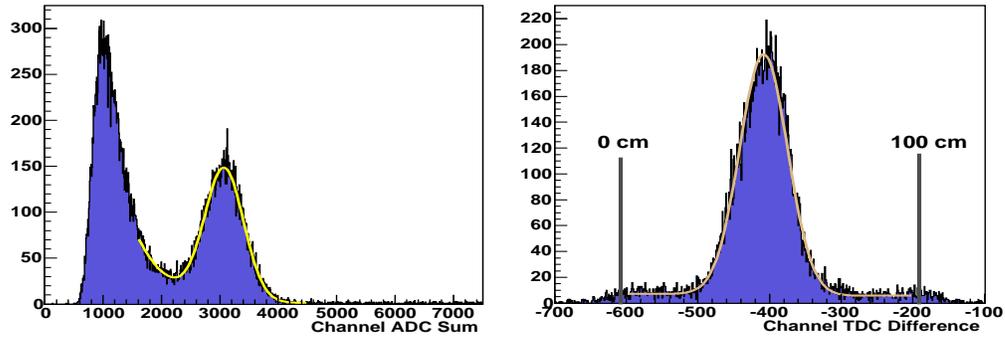}\end{center}

\caption{{\small \label{fig:In_ADC-TDC}CBS spectra for the prototype cell
loaded with the In-h2 sample, $^{137}\textrm{Cs}$ source at cell
center.}}

{\small Left: distribution of the total charge, calculated by summing
the ADC values of the two PMTs.}{\small \par}

{\small Right: distribution of the time difference between the two
PMTs. The TDC signals corresponding to the ends of the cell are indicated
by the two vertical segments.}
\end{figure}

Fig. \ref{fig:In_ADC-TDC} shows the ADC and TDC spectra of a high
statistics CBS run with the In-h2 sample.%
The analysis of the CBS runs is reported in Table
\ref{tab:In-high_CBS-analysis}. %
At $44\,\textrm{g/l}$ In-loading, $\sigma_{E}=(11.6\pm0.2)\%$ and
$\sigma_{x}=(7.0\pm0.4)\,\textrm{cm}$ at $477\,\textrm{keV}$, source
at center. The results of In-h1 and In-h2 are compatible with each other
within the errors.

\subsubsection{Photo-electron yield}

The estimations of the PY are given in Table \ref{tab:In_PY} %
 The values derived from the \emph{spe} calibrations are in this case
significantly higher than those based on the energy resolution, probably
because for a low number of \emph{pe} the departure of the energy
resolution from the $\frac{1}{\sqrt{N}}$ statistical limit is more
important. The light at the cell center is for all In-loaded samples
less than 1/3 of PXE. The difference of LY gives a factor $\sim1/2$
and the lower transparency explains the rest. In fact, the MC predicts
a similar ratio as observed.

\subsubsection{Long-term stability}

We have re-measured the optical properties of the prototype cell $\sim14$
months after the first campaign (sample In-h2). During this time the
cell has been stored in darkness, at room temperature (no temperature
control) and exposed to air (sealed in nitrogen atmosphere, however
the air-tightness of the module was not tested). The results of light
attenuation, energy resolution and PY are presented in Table
\ref{tab:summary-14months-later}. 
The results are overall consistent with our first determination
(Tables \ref{tab:In_att-fitpar}, \ref{tab:In-high_CBS-analysis}
and \ref{tab:In_PY}). The large error of the long-range attenuation
length is due to the fact that less points have been probed, however
for all measured positions the ratio $I(x)/I(50\, cm)$ is well consistent
with the results of Fig. \ref{fig:In-high_att}. The energy resolutions
of the single PMTs and the \emph{spe} calibration indicate a slightly
higher light output, which we however consider not significant, taking
into account the systematic uncertainties related to the re-building
of the electronics and the new optical coupling of the PMTs (not included
in the quoted errors). We conclude that neither the transparency,
nor the LY of the scintillator have degraded during a period of $>1\,\textrm{y}$.

\subsection{\label{sub:Summary-and-Discussion}Summary and discussion}

\begin{figure}
\begin{center}\includegraphics[%
  width=1.0\columnwidth,
  keepaspectratio]{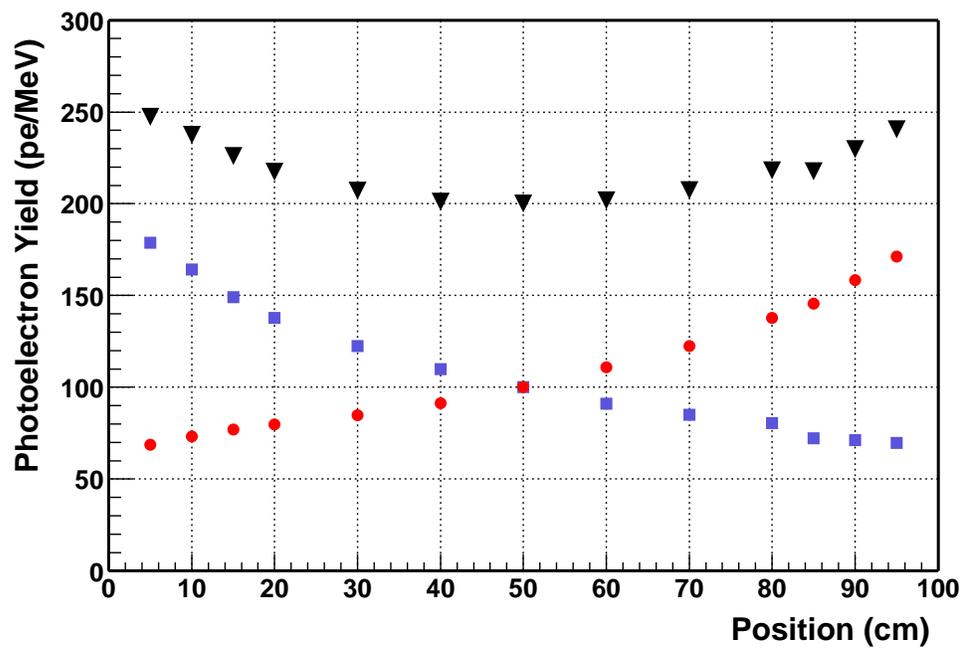}\end{center}

\caption{{\small \label{fig:In44_summary-PY}PY as a function of the source
position for the In-h2 prototype cell. The response of both PMTs (square
and circular markers) and the total charge (triangular markers) are
shown.}}
\end{figure}

The measurements of PY and light attenuation are brought together
in Fig. \ref{fig:In44_summary-PY}, which shows the absolute detector
performance as a function of the source position for the In-h2 sample
(which has a similar composition as the sample we filled in the Gran
Sasso pilot detector). %

\begin{figure}
\begin{center}\includegraphics[%
  width=1.0\columnwidth]{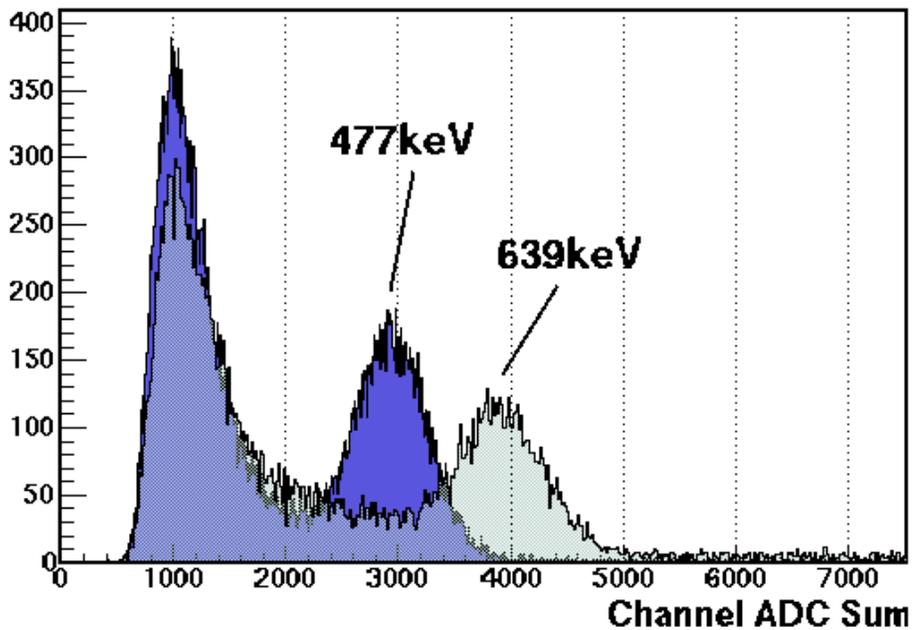}\end{center}

\caption{{\small \label{fig:In44_Cs-Mn}Resolution of two spectral lines with
the MPIK prototype, In-h1 sample.}}
\end{figure}

In Fig. \ref{fig:In44_Cs-Mn} the CBS spectra of two runs of similar
statistics are superimposed, one with the $^{137}$Cs source, the
other with $^{54}$Mn (In-h1 sample). The figure gives an idea of
how well the MPIK LENS prototype would energetically resolve an event
in the upper range of the $^{115}\textrm{In}$ $\beta$-background
($\textrm{E}\lesssim500\,\textrm{keV}$) from the delayed $\nu$-tag
$(\textrm{E$=613\,$ keV})$. %
The resolution of these two events is the crucial parameter for the
feasibility of an In solar neutrino experiment. With the results presented
in this paper, at $44\,\textrm{g/l}$ In-loading those two energies
are $\sim1.5\,\sigma$ apart, where $\sigma$ is given by the sum
in quadrature of the standard deviation of the single lines. For example,
if a cut is set at the $\nu$-tag energy minus $1\sigma$, the probability
that a background event with $\textrm{E}\sim480\,\textrm{keV}$ survives
the cut, considering only the deposited energy, is $\sim10\,\%$.
A detailed Monte Carlo analysis of the signal to noise ratio has been
carried out within the LENS collaboration \cite{Meyer}. It has shown
that the suppression of the In $\beta$-Bremsstrahlung background
to an adequate level requires an energy resolution at $\sim480\,\textrm{keV}$
of a factor of $\sim30\,\%$ better than that reported here.

\begin{figure}
\begin{center}\includegraphics[%
  width=1.0\columnwidth,
  keepaspectratio]{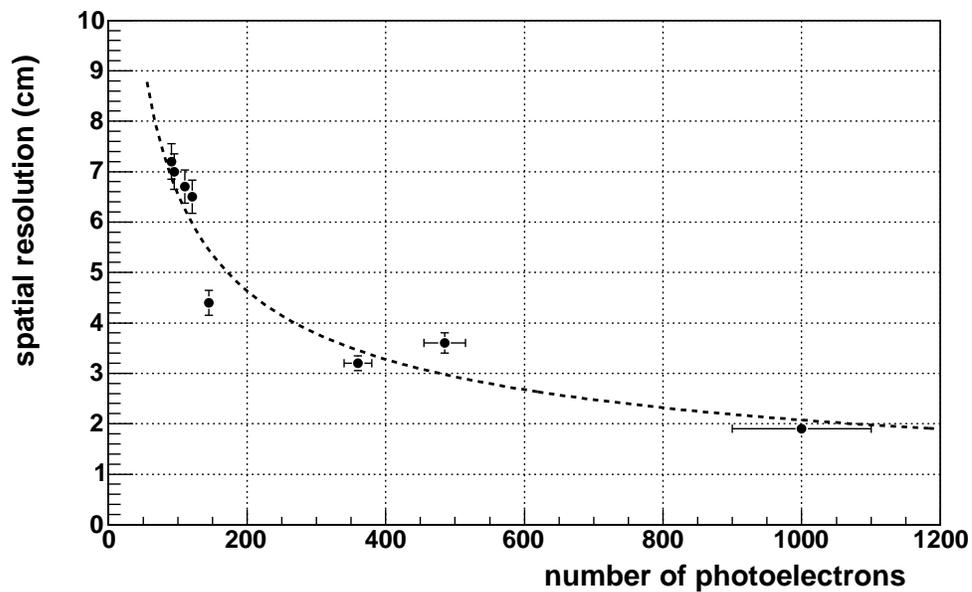}\end{center}

\caption{{\small \label{fig:spres-vs-Npe}Experimental spatial resolution
as a function of the total number of} \emph{\small pe}{\small . Each
marker represents one of the measurements reported in this paper.
The dashed line is a fit with the function $f(N_{pe})=\frac{65.55}{\sqrt{N_{pe}}}$.}}
\end{figure}

In Fig. \ref{fig:spres-vs-Npe} all the measurements of spatial resolution
have been displayed as a function of the number of \emph{pe}. %
The spatial resolution can be parametrized by a function $\sigma_{x}\sim\frac{k}{\sqrt{N_{pe}}}$,
where $N_{pe}$ is the number of \emph{pe} and $k$ a constant. The
fit gives $k\sim65.6$. The scatter of the data with respect to the
fit function is large, however the parametrization gives a useful
{}``rule of thumb'' to predict the detector spatial resolution for
a wide range of cases with better than $20\,\%$ precision.

The last column of Tables \ref{tab:PXE_PY} and \ref{tab:In_PY} reminds
us that more than a factor 2 of light is missing according to the
MC predictions. The fact that the simulations reproduce reasonably
well the experimental light attenuation curves demonstrates that the
discrepancy in the PY is not related to light losses {}``on the way''
to the PMTs. This is supported by the observation that the experimental
deficit relative to the MC is roughly constant. It is possible that the
$2^{\prime\prime}$ PMTs utilized have an effective photocathode coverage
smaller than the specification, or that the PMTs are less sensitive
in a broad outer photocathode area, compared to the center. This explanation
is corroborated by the results of the LNGS-INR group of the LENS collaboration,
which has measured at Gran Sasso an identical quartz prototype cell
filled with an In hydroxy-carboxylate scintillator \cite{LNGS_scint}.
They have used $3^{\prime\prime}$ PMTs, which ensure full coverage
of the cell ends. Their reported results \cite{LNGS-cell-optics} are in
agreement with our MC expectations.

\section{\label{sec:Conclusions-Outlook}Conclusions and outlook}

In this paper we have described the work carried out at MPIK to design,
model, build and characterize a prototype cell of an indium solar
neutrino detector filled with a novel In-loaded scintillator developed
by us (In-acac).

As a first step toward the validation of the experimental concept,
we have experimentally demonstrated that light-piping via total internal
reflection on quartz and specular reflection by 3M-VM2000 mirror foil
is adequate for a fine segmented detector. 

We have then predicted by Monte Carlo the expected optical performances
of an In-acac cell as a function of the scintillator composition,
in order to optimize the formulation at any given indium concentration. 

After this preparatory phase, we have set up a prototype cell and
a system to measure the detector optical performances. A benchmark
PXE-based scintillator has shown an effective long-range attenuation
length of $\sim4.2\,\textrm{m}$ and a photo-electron yield of $\sim730\, pe/\textrm{MeV}$.
These results give the experimental upper limit of the potential of
an In-based solar neutrino detector in the considered implementation. 

Subsequently, three different In-loaded samples have been measured.
At an indium concentration of $44\,\textrm{g/l}$, the energy resolutions
was $\sim11.6\,\%$ and the spatial resolution $\sim7\,\textrm{cm}$,
both for $477\,\textrm{keV}$ recoil electrons.
The long-range attenuation length in the cell was
$\sim1.3\,\textrm{m}$ and the photo-electron yield $\sim200\, pe/\textrm{MeV}$.

For all samples, the measured light attenuation was in good agreement
with the prediction of our MC simulation, and the light output as
well, after normalization by an experimental quality factor of $\sim0.45$.
This means that we can reliably predict the performance of this detector
for any choice of the scintillator composition.

For our base-case In-loading of $44\,\textrm{g/l}$, we found that
the energy separation between the In $\beta$ end-point and the $\nu$-signature
is of the order of $\sim1.5\,\sigma$. According to Monte Carlo studies
of the full-scale detector, this is not yet sufficient to guarantee
an adequate signal/background ratio. However, we believe that there
is still room to increase the light output by simply improving the match
of the PMT surface to the cell end. Unfortunately, the attenuation length we
measured (which is fluor-dominated) would limit the length of the
unitary cell to $\lesssim1.5\,\textrm{m}$.

The long-term stability of any metal-loaded scintillator is one of
the crucial issues for the feasibility of a neutrino detector. We
have demonstrated the our system is stable and robust. This scintillator
did not show any degradation of its optical performance for a period
of $>1\,\textrm{y}$, in spite of the very high indium concentration
and storage in an unprotected environment. This result is of great
interest not only for a solar neutrino detector, but also for loading
of Gd in a novel liquid scintillator, to use in upcoming reactor anti-neutrino
detectors \cite{Double-Chooz}.

A prototype cell filled with a scintillator similar to the In-h2 sample
reported here has been installed in the LLBF, as a part of the LENS
pilot experiment at Gran Sasso \cite{LENS-Nu2004}. The background
above the $^{115}$In $\beta$ end-point was measured and the rate
resulted only of a factor of $\sim2$ higher than in high-purity benchmark
PXE cells.

\section*{Acknowledgments}

We would like to thank the technical support of our mechanical and
glassblowing workshops, especially F. Kleinbongardt, P. M\"ogel and
E. Borger for their help in the realization of the light piping measurements
and of the prototype set-up. The contribution of notable amounts of
BPO fluor from our Russian colleagues, Dr. Leonid Bezrukov and group
is acknowledged. We also thank all the members of the LENS collaboration
for stimulating and helpful discussions. C. Buck thanks the Graduiertenkolleg,
Ruprecht-Karls-Universit\"at Heidelberg, for fellowship support.

\hfill
\newpage

\hfill
\newpage

\hfill
\newpage

\hfill
\newpage

\hfill
\newpage

\hfill
\newpage

\hfill
\newpage

\hfill
\newpage

\hfill
\newpage

\hfill
\newpage

\hfill
\newpage

\hfill
\newpage

\hfill
\newpage

\hfill
\newpage

\hfill
\newpage

\hfill
\newpage

\begin{table}

\caption{{\small \label{tab:fluoresc-param}Estimation of the probability
$\phi$ that an absorption results in a new photon emission for all
detector components. }}

\begin{center}{\small }\begin{tabular}{c||c|c|c|c|c}
&
{\small Anisol}&
{\small In(acac)$_{3}$}&
{\small BPO}&
{\small bis-MSB}&
{\small Quartz }\tabularnewline
\hline
\hline 
{\small $\phi$}&
 {\small $0\,\%\forall\lambda$}&
 {\small $0\,\%\forall\lambda$}&
{\small $\begin{array}{l}
90\,\%\:\lambda<410\,\textrm{nm}\\
0\,\%\:\lambda>410\,\textrm{nm}\end{array}$}&
{\small $\begin{array}{l}
94\,\%\:\lambda<440\,\textrm{nm}\\
0\,\%\:\lambda>440\,\textrm{nm}\end{array}$}&
 {\small $0\,\%\forall\lambda$}\tabularnewline
\end{tabular}\end{center}
\end{table}

\hfill
\newpage

\begin{table}

\caption{{\small \label{tab:samples}Samples measured in the prototype cell.
Fluor is p-TP (para-terphenyl, 1,4-diphenyl-benzol) for PXE, BPO for
the In-loaded scintillators. WLS is bis-MSB. The LY is given in percent
of BC505 and has a relative error of $\sim5\,\%$.}}

\begin{center}{\small }\begin{tabular}{c|c|c|c|c|c}
{\small Name}&
{\small solvent}&
{\small In}&
{\small Fluor}&
{\small WLS}&
{\small LY}\tabularnewline
&
&
{\small (g/l)}&
{\small (g/l)}&
{\small (mg/l)}&
{\small (\%)}\tabularnewline
\hline
\hline 
{\small PXE}&
{\small PXE}&
{\small 0}&
{\small 2 }&
{\small 20}&
{\small 83}\tabularnewline
\hline 
{\small In-l}&
{\small anisol}&
{\small 20}&
{\small 27}&
{\small 50}&
{\small 44}\tabularnewline
\hline 
{\small In-h1}&
{\small anisol}&
{\small 44}&
{\small 47}&
{\small 50}&
{\small 42}\tabularnewline
\hline 
{\small In-h2}&
{\small anisol}&
{\small 44}&
{\small 47}&
{\small 100}&
{\small 42}\tabularnewline
\end{tabular}\end{center}
\end{table}

\hfill
\newpage

\begin{table}

\caption{{\small \label{tab:samples_att-lengths}Attenuation lengths at various
wavelengths for the measured scintillator samples.}}

\begin{center}{\small }\begin{tabular}{cc|c|c|c|c|c|c|c|c}
{\small $\lambda$ (nm)}&
&
{\small 380}&
{\small 400}&
{\small 410}&
{\small 420}&
{\small 430}&
{\small 440}&
{\small 450}&
{\small 500}\tabularnewline
\hline
\hline 
{\small $\mu$ (cm)}&
{\small PXE}&
{\small 0.25}&
{\small 5.2}&
{\small 78}&
{\small $5.8\,10^{2}$}&
{\small $1.3\,10^{3}$}&
{\small $1.7\,10^{3}$}&
{\small $2.3\,10^{3}$}&
{\small $9.5\,10^{3}$}\tabularnewline
\hline 
{\small $\mu$ (cm)}&
{\small In-l}&
{\small 0.09}&
{\small 0.99}&
{\small 13}&
{\small 76}&
{\small $1.4\,10^{2}$}&
{\small $1.7\,10^{2}$}&
{\small $2.0\,10^{2}$}&
{\small $5.4\,10^{2}$}\tabularnewline
\hline 
{\small $\mu$ (cm)}&
{\small In-h1}&
{\small 0.08}&
{\small 0.93}&
{\small 12}&
{\small 55}&
{\small 88}&
{\small $1.1\,10^{2}$}&
{\small $1.3\,10^{2}$}&
{\small $3.4\,10^{2}$}\tabularnewline
\hline 
{\small $\mu$ (cm)}&
{\small In-h2}&
{\small 0.05}&
{\small 0.50}&
{\small 6.6}&
{\small 43}&
{\small 83}&
{\small $1.1\,10^{2}$}&
{\small $1.3\,10^{2}$}&
{\small $3.4\,10^{2}$}\tabularnewline
\end{tabular}\end{center}
\end{table}

\hfill
\newpage

\begin{table}

\caption{{\small \label{tab:pxe_att-fitpar}Parameters returned by a double
exponential fit to the experimental data and the simulations of Figs.
\ref{fig:pxe_tir_att} and \ref{fig:pxe_foils_att}. The fit of the
experimental data is reported after averaging the response of the
PMTs. $\dag$: {}``TIR or SR'', $\S$: {}``TIR and SR''. See text
for the definitions. The values in brackets indicate the $1\sigma$
error in the units of the least significant digit.}}

\begin{center}{\small }\begin{tabular}{c|c|c|c|c}
{\small source}&
{\small piping}&
{\small $\mu_{s}$}&
{\small $\mu_{l}$}&
{\small $W_{l}$}\tabularnewline
&
&
{\small cm}&
{\small cm}&
\tabularnewline
\hline
\hline 
{\small LED}&
{\small TIR}&
{\small $11.2$}&
{\small $288(5)$}&
{\small $0.888(5)$}\tabularnewline
\hline 
{\small $^{137}$Cs}&
{\small TIR}&
{\small $24.6$}&
{\small $305(7)$}&
{\small $0.849(5)$}\tabularnewline
\hline 
{\small MC}&
{\small TIR}&
{\small $13.6$}&
{\small $330(16)$}&
{\small $0.819(8)$}\tabularnewline
\hline
\hline 
{\small $^{137}$Cs}&
{\small TIR/SR}&
{\small $16.2$}&
{\small $418(29)$}&
{\small $0.906(10)$}\tabularnewline
\hline 
{\small MC}&
{\small TIR/SR$^{\dag}$}&
{\small $16.6$}&
{\small $316(7)$}&
{\small $0.921(4)$}\tabularnewline
\hline 
{\small MC}&
{\small TIR/SR$^{\S}$}&
{\small $14.8$}&
{\small $375(13)$}&
{\small $0.950(7)$}\tabularnewline
\end{tabular}\end{center}
\end{table}

\hfill
\newpage

\begin{table}

\caption{{\small \label{tab:pxe_CBS-analysis}Analysis of the CBS measurements
with PXE. $\sigma_{E}$ and $\sigma_{x}$ are energy and spatial resolution;
the labels} \emph{\small l}{\small ,} \emph{\small r} {\small and}
\emph{\small s} {\small stand for {}``left'', {}``right'' and
{}``sum'', respectively. The position {}``x'' is measured from
the {}``left'' PMT. The values in brackets indicate the $1\sigma$
error in the units of the least significant digit.}}

\begin{center}{\small }\begin{tabular}{c|c|c|c|c|c|c}
{\small En}&
{\small piping}&
x&
{\small $\sigma_{E}^{l}$ }&
{\small $\sigma_{E}^{r}$}&
{\small $\sigma_{E}^{s}$ }&
{\small $\sigma_{x}$}\tabularnewline
{\small (keV)}&
&
{\small (cm)}&
{\small (\%)}&
{\small (\%)}&
{\small (\%)}&
{\small (cm)}\tabularnewline
\hline
\hline 
{\small 477}&
{\small TIR}&
{\small 10}&
{\small $10.8(2)$}&
{\small $13.5(3)$}&
{\small $8.7(2)$}&
{\small $4.9(3)$}\tabularnewline
\hline 
{\small 477}&
{\small TIR}&
{\small 30}&
{\small $11.6(2)$}&
{\small $12.9(3)$}&
{\small $8.8(2)$}&
{\small $5.0(3)$}\tabularnewline
\hline
{\small 477}&
{\small TIR}&
{\small 50}&
{\small $13.2(3)$}&
{\small $11.9(2)$}&
{\small $9.0(2)$}&
{\small $4.4(2)$}\tabularnewline
\hline
{\small 477}&
{\small TIR}&
{\small 70}&
{\small $13.4(3)$}&
{\small $11.1(2)$}&
{\small $8.7(2)$}&
{\small $4.1(3)$}\tabularnewline
\hline
\hline 
{\small 477}&
{\small TIR/SR}&
{\small 10}&
{\small $7.1(2)$}&
{\small $7.3(2)$}&
{\small $5.5(2)$}&
{\small $4.1(2)$}\tabularnewline
\hline
{\small 477}&
{\small TIR/SR}&
{\small 30}&
{\small $7.5(2)$}&
{\small $7.0(2)$}&
{\small $5.3(2)$}&
{\small $4.0(2)$}\tabularnewline
\hline
{\small 477}&
{\small TIR/SR}&
{\small 50}&
{\small $7.6(2)$}&
{\small $6.8(2)$}&
{\small $5.4(2)$}&
{\small $3.2(2)$}\tabularnewline
\hline
{\small 639}&
{\small TIR/SR}&
{\small 50}&
{\small $6.6(2)$}&
{\small $5.5(2)$}&
{\small $4.6(2)$}&
{\small $3.6(2)$}\tabularnewline
\end{tabular}\end{center}
\end{table}

\hfill
\newpage

\begin{table}

\caption{{\small \label{tab:PXE_PY}Estimations of the PY for the PXE sample.
Source at the cell center. The column of the calibration reports the
value estimated multiplying by 2 the average number of} \emph{\small pe}
{\small detected by the better calibrated {}``right'' PMT. The two
limits in the range indicated for the MC in the second row correspond
to the cases {}``TIR or SR'' and {}``TIR and SR'' discussed in
the text. For the ratio in the last column, Exp is the the higher
PY estimation between resolution and calibration.}}

\begin{center}{\small }\begin{tabular}{c|c|c|c|c}
{\small piping}&
{\small resolution}&
{\small calibration}&
{\small MC }&
{\small $\underline{\textrm{Exp}}$}\tabularnewline
&
{\small (}\emph{\small pe}{\small /MeV)}&
{\small (}\emph{\small pe}{\small /MeV)}&
{\small (}\emph{\small pe}{\small /MeV)}&
\emph{\small MC}\tabularnewline
\hline
\hline 
{\small TIR}&
{\small $\gtrsim260$}&
{\small $305\pm15$}&
{\small $\sim680$}&
{\small $\sim0.45$}\tabularnewline
\hline 
{\small TIR/SR}&
{\small $\gtrsim730$}&
{\small $715\pm30$}&
{\small $\sim1570-1730$}&
{\small $\sim0.42-0.46$}\tabularnewline
\end{tabular}\end{center}
\end{table}

\hfill
\newpage

\begin{table}

\caption{{\small \label{tab:In_att-fitpar}Parameters returned by a double
exponential fit to the experimental and simulated light attenuation
curves. The values in brackets indicate the $1\sigma$ error in the
units of the least significant digit.}}

\begin{center}{\small }\begin{tabular}{c|c|c|c}
&
{\small $\mu_{s}$}&
{\small $\mu_{l}$}&
{\small $W_{l}$}\tabularnewline
&
{\small (cm)}&
{\small (cm)}&
\tabularnewline
\hline
\hline 
{\small In-l}&
{\small $23.1$}&
{\small $154(17)$}&
{\small $0.72(3)$}\tabularnewline
\hline 
{\small In-h1}&
{\small $17.7$}&
{\small $127(5)$}&
{\small $0.75(1)$}\tabularnewline
\hline 
{\small In-h2}&
{\small $16.4$}&
{\small $131(9)$}&
{\small $0.79(2)$}\tabularnewline
\hline
\hline 
{\small MC In-l }&
{\small $16.5$}&
{\small $161(3)$}&
{\small $0.798(6)$}\tabularnewline
\hline 
{\small MC In-h1}&
{\small $20.7$}&
{\small $140(5)$}&
{\small $0.663(9)$}\tabularnewline
\hline 
{\small MC In-h2}&
{\small $19.8$}&
{\small $138(5)$}&
{\small $0.675(9)$}\tabularnewline
\end{tabular}\end{center}
\end{table}

\hfill
\newpage

\begin{table}

\caption{{\small \label{tab:In-high_CBS-analysis}Analysis of the CBS measurements
of the In-loaded samples. Source at the cell center. The values in
brackets indicate the $1\sigma$ error in the units of the least significant
digit.}}

\begin{center}{\small }\begin{tabular}{c|c|c|c|c|c}
{\small Sample}&
{\small En}&
{\small $\sigma_{E}^{l}$}&
{\small $\sigma_{E}^{r}$}&
{\small $\sigma_{E}^{s}$}&
{\small $\sigma_{x}$}\tabularnewline
&
{\small (keV)}&
{\small (\%)}&
{\small (\%)}&
{\small (\%)}&
{\small (}\emph{\small cm}{\small )}\tabularnewline
\hline
\hline 
{\small In-l}&
{\small 477}&
{\small $14.5(6)$}&
{\small $13.8(4)$}&
{\small $10.8(2)$}&
{\small $6.7(3)$}\tabularnewline
\hline
{\small In-h1}&
{\small 477}&
{\small $18.4(2)$}&
{\small $15.6(2)$ }&
{\small $11.8(2)$}&
{\small $7.2(4)$}\tabularnewline
\hline
{\small In-h1}&
{\small 639}&
{\small $14.4(2)$}&
{\small $12.8(2)$}&
{\small $10.1(2)$}&
{\small $6.5(3)$}\tabularnewline
\hline
{\small In-h2}&
{\small 477}&
{\small $18.6(3)$}&
{\small $15.3(2)$}&
{\small $11.6(2)$}&
{\small $7.0(4)$}\tabularnewline
\end{tabular}\end{center}
\end{table}

\hfill
\newpage

\begin{table}

\caption{{\small \label{tab:In_PY}Estimations of the PY for the measurement
of the In-loaded samples, with source at the cell center.}}

\begin{center}{\small }\begin{tabular}{c|c|c|c|c}
{\small Sample}&
{\small resolution}&
{\small calibration}&
{\small MC }&
{\small $\underline{\textrm{Exp}}$}\tabularnewline
&
{\small (}\emph{\small pe}{\small /MeV)}&
{\small (}\emph{\small pe}{\small /MeV)}&
{\small (}\emph{\small pe}{\small /MeV)}&
\emph{\small MC}\tabularnewline
\hline
\hline 
{\small In-l}&
{\small $\gtrsim180$}&
{\small $230(10)$}&
{\small $\sim565$}&
{\small $\sim0.41$}\tabularnewline
\hline 
{\small In-h1}&
{\small $\gtrsim150$}&
{\small $190(10)$}&
{\small $\sim395$}&
{\small $\sim0.48$}\tabularnewline
\hline 
{\small In-h2}&
{\small $\gtrsim155$}&
{\small $200(10)$}&
{\small $\sim400$}&
{\small $\sim0.50$}\tabularnewline
\end{tabular}\end{center}
\end{table}

\hfill
\newpage

\begin{table}

\caption{{\small \label{tab:summary-14months-later}Summary of the optical
measurements performed on the cell filled with the In-h2 sample, 14
months after the first campaign. The second rows reiterates for reference
the results reported in Tables \ref{tab:In_att-fitpar}, \ref{tab:In-high_CBS-analysis}
and \ref{tab:In_PY}. All resolutions refer to an energy deposition
of $477\,\textrm{keV}$. The values in brackets indicate the $1\sigma$
error in the units of the least significant digit.}}

\begin{center}{\small }\begin{tabular}{c|c|c|c|c|c|c|c|c}
{\small Date}&
{\small $\mu_{s}$}&
{\small $\mu_{l}$}&
{\small $W_{l}$}&
{\small $\sigma_{E}^{l}$}&
{\small $\sigma_{E}^{r}$}&
{\small $\sigma_{E}^{s}$}&
{\small $\sigma_{x}$}&
{\small calibration}\tabularnewline
&
{\small (cm)}&
{\small (cm)}&
&
{\small (\%)}&
{\small (\%)}&
{\small (\%)}&
{\small (}\emph{\small cm}{\small )}&
{\small (}\emph{\small pe}{\small /MeV)}\tabularnewline
\hline
\hline 
{\small Oct 04}&
{\small $18.5$}&
{\small $138(20)$}&
{\small $0.73(4)$}&
{\small $17.3(5)$}&
{\small $14.8(3)$}&
{\small $11.7(2)$}&
{\small $6.9(4)$}&
{\small $230\pm10$}\tabularnewline
\hline 
{\small Aug 03}&
{\small $16.4$}&
{\small $131(9)$}&
{\small $0.79(2)$}&
{\small $18.6(3)$}&
{\small $15.3(2)$}&
{\small $11.6(2)$}&
{\small $7.0(4)$}&
{\small $200\pm10$}\tabularnewline
\end{tabular}\end{center}
\end{table}

\hfill

\end{document}